\documentclass[twocolumn,aps,pra,superscriptaddress,showpacs,nofootinbib]{revtex4-1}
\usepackage{epsfig}
\usepackage[colorlinks=true,citecolor=blue,urlcolor=blue,linkcolor=blue]{hyperref}
\usepackage{chemscheme}
\usepackage{graphicx}
\usepackage{multirow}
\usepackage[table,xcdraw]{xcolor}
\usepackage{natbib}
\usepackage{chemfig}
\usepackage[version=4]{mhchem}
\usepackage[normalem]{ulem}
\usepackage{floatrow}
\usepackage{tabulary}
\usepackage{booktabs}
\usepackage[flushleft]{threeparttable}
\usepackage{xspace}
\DeclareMathAlphabet{\mathbsf}{OT1}{cmss}{bx}{n}

\newcommand{\cm}{cm$^{-1}$\xspace}
\newcommand{\A}{$\widetilde A$\xspace}
\newcommand{\X}{$\widetilde X$\xspace}
\newcommand{\B}{$\widetilde B$\xspace}

\usepackage{tablefootnote}

\begin{document}

\title{Bottom-up approach to scalable growth of molecules capable of optical cycling}

\affiliation{Department of Physics and Astronomy, University of California, Los Angeles, California 90095, USA}
\affiliation{Department of Chemistry and Biochemistry, University of California, Los Angeles, California 90095, USA}
\affiliation{Institut f\"ur Chemie, Humboldt-Universit\"at zu Berlin, Berlin 10099, Germany}
\author{Guanming Lao}
\affiliation{Department of Physics and Astronomy, University of California, Los Angeles, California 90095, USA}
\affiliation{These authors contributed equally.}

\author{Taras Khvorost}
\affiliation{Department of Chemistry and Biochemistry, University of California, Los Angeles, California 90095, USA}
\affiliation{These authors contributed equally.}

\author{Antonio Macias, Jr.}
\affiliation{Department of Chemistry and Biochemistry, University of California, Los Angeles, California 90095, USA}

\author{Harry W. T. Morgan}
\affiliation{Department of Chemistry and Biochemistry, University of California, Los Angeles, California 90095, USA}

\author{Robert H. Lavroff}
\affiliation{Department of Chemistry and Biochemistry, University of California, Los Angeles, California 90095, USA}

\author{Ryan Choi}
\affiliation{Department of Chemistry and Biochemistry, University of California, Los Angeles, California 90095, USA}

\author{Haowen Zhou}
\affiliation{Department of Physics and Astronomy, University of California, Los Angeles, California 90095, USA}

\author{Denis Usvyat}
\affiliation{Institut f\"ur Chemie, Humboldt-Universit\"at zu Berlin, Berlin 10099, Germany}

\author{Guo-Zhu Zhu}
\thanks{Present address: Department of Chemistry, Fudan University, Shanghai 200438, China}
\affiliation{Department of Physics and Astronomy, University of California, Los Angeles, California 90095, USA}
\email{gzhu2019@g.ucla.edu}
\affiliation{Corresponding author}

\author{Miguel A. García-Garibay}
\affiliation{Department of Chemistry and Biochemistry, University of California, Los Angeles, California 90095, USA}
\email{mgg@chem.ucla.edu}
\affiliation{Corresponding author}

\author{Anastassia N. Alexandrova}
\affiliation{Department of Chemistry and Biochemistry, University of California, Los Angeles, California 90095, USA}
\email{ana@chem.ucla.edu}
\affiliation{Corresponding author}

\author{Eric R. Hudson}
\affiliation{Department of Physics and Astronomy, University of California, Los Angeles, California 90095, USA}
\email{eric.hudson@ucla.edu}
\affiliation{Corresponding author}

\author{Wesley C. Campbell}
\affiliation{Department of Physics and Astronomy, University of California, Los Angeles, California 90095, USA}
\affiliation{Corresponding author}
\email{wes@physics.ucla.edu}

\date{\today}

\begin{abstract}

Gas-phase molecules capable of repeatable, narrow-band spontaneous photon scattering are prized for direct laser cooling and quantum state detection.  Recently, large molecules incorporating phenyl rings have been shown to exhibit similar vibrational closure to the small molecules demonstrated so far, and it is not yet known if the high vibrational-mode density of even larger species will eventually compromise optical cycling. Here, we systematically increase the size of hydrocarbon ligands attached to single alkaline-earth-phenoxides from (-H) to -C$_{14}$H$_{19}$ while measuring the vibrational branching fractions of the optical transition.  We find that varying the ligand size from 1 to more than 30 atoms
does not systematically reduce the cycle closure, which remains around 90\%. Theoretical extensions to larger diamondoids and bluk diamond surface suggest that alkaline earth phenoxides may maintain the desirable scattering behavior as the system size grows further, with no indication of an upper limit. 
\end{abstract}

\maketitle

\section*{Introduction }
Quantum information processing (QIP) has been explored in a wide variety of physical platforms, such as superconducting circuits \cite{krantz2019superconducting}, semiconductor spins \cite{zwanenburg2013silicon}, ultracold neutral atoms \cite{kaufman2021quantum} and trapped atomic ions \cite{bruzewicz2019trapped}.

As a QIP platform, molecules offer some unique advantages, including the potential for atomic-scale customization through chemical assembly, laboratory-controllable electric dipole moments for fast gates, and a variety of internal degrees of freedom for encoding quantum information \cite{carr2009cold}. In particular, molecules containing optical cycling centers (OCCs) \cite{Isaev2016Polyatomic,kozyryev2016proposal,augenbraun2020molecular,Dickerson2021FranckCondon}, which allow repeated, state-dependent absorption and emission of photons, enable direct laser cooling \cite{fitch2021laser,vilas2022magneto,augenbraun2023direct} and high-fidelity quantum state detection \cite{cheuk2018lambda,shaw2021resonance}, making them promising as hosts for high-quality qubits.

One way to access the features furnished by molecules for quantum information processing is to load gas-phase molecules hosting qubits into optical tweezer arrays \cite{schlosser2001sub,kaufman2021quantum}. This approach has been successfully demonstrated with 
laser-cooled diatomic molecules \cite{anderegg2019optical,holland2023bichromatic,vilas2024optical}, whose electric dipole-dipole interactions have recently been used to create entanglement \cite{Holland2023OnDemand,Bao2023Dipolar}.  State preparation and measurement (SPAM) of these molecular qubits is achieved via optical cycling \cite{cheuk2018lambda}, which can be done quickly and non-destructively and is therefore also attractive for processors based on trapped molecular ions \cite{Campbell2020DipolePhonon,Hudson2021Laserless}.  Extensions of these ideas to larger (polyatomic) molecules have been proposed to take advantage of the additional flexibility afforded by those species \cite{yu2019scalable}, and a tweezer array of optical-cycling triatomic molecules (CaOH) has been demonstrated \cite{vilas2024optical}.  In particular, larger molecular species can furnish smaller splittings between opposite-parity energy eigenstates, which allows them to be oriented in the laboratory frame with only modest fields \cite{Kozyryev2017Precision}.  Further, the existence of accessible manifolds of large total angular momentum is necessary to realize recently-proposed schemes for robust encoding of quantum information \cite{Albert2020Robust,jain2023ae}.

An alternative platform that provides access to the advantages provided by molecular qubits is to tether OCC-bearing molecules hosting qubits to a substrate \textit{in vacuo} to form lithographically-definable surface-bound qubit arrays \cite{guo2021surface}.  Much like defect centers in solids \cite{wolfowicz2021quantum} (such as the negative nitrogen-vacancy center in diamond), this removes the need to trap gas-phase species,
and the resulting simplification of the necessary apparatus is attractive for scaling to large processors.  This approach can in some ways be regarded as the large-ligand limit of the gas-phase platform, where the common feature of the two is the OCC that enables state preparation and measurement (SPAM).

Finding molecules that are capable of optical cycling, however, is challenging. Molecular optical cycling and direct laser cooling were first demonstrated for 
diatomic species,~\cite{shuman2010laser,hummon20132d,zhelyazkova2014laser,anderegg2017radio,lim2018laser,albrecht2020buffer,zhang2022doppler,gu2022radiative,hofsass2021optical} and have now been extended to some triatomic ~\cite{kozyryev2017sisyphus,baum20201d,augenbraun2020laser} and a small polyatomic molecule CaOCH$_3$  ~\cite{mitra2020direct}.
Spectroscopic investigations of Ca-containing and Sr-containing phenoxides along with their derivatives via molecular functionalization, \cite{zhu2022functionalizing,lao2022sroph,mitra2022pathway,augenbraun2022high} have shown promise for optical cycling of much larger species with measured vibrational branching fractions of 0.84-0.99.  Further, a variety of large polyatomic OCC systems have been proposed theoretically for these applications, including alkaline-earth-metal-containing alkoxides \cite{Isaev2016Polyatomic,kozyryev2016proposal,dickerson2022fully,augenbraun2020molecular}, arenes
\cite{ivanov2019towards,ivanov2020two,ivanov2020toward,Dickerson2021FranckCondon,dickerson2021optical}, fullerenes \cite{klos2020prospects}, diamondoids \cite{dickerson2022fully}, and surfaces \cite{guo2021surface}.

As the size of the molecular ligand to which an OCC is attached grows, however, challenges can arise from the complexity this introduces.  For example,  intricate vibrational coupling, such as Fermi resonance, can introduce additional vibrational decay pathways that require extra repumping lasers \cite{Zhang2023IntensityBorrowing,zhu2024extending}.  This raises questions about the viability of optical cycling centers beyond molecules containing just a few atoms. Can the optical cycling properties of functionalized molecules be maintained as we scale up the molecular size? Do surface-bound OCCs perform comparably to their gas-phase counterparts, thereby offering a stable QIS platform?

\begin{scheme*}
    \centering
    \includegraphics[scale=0.8]{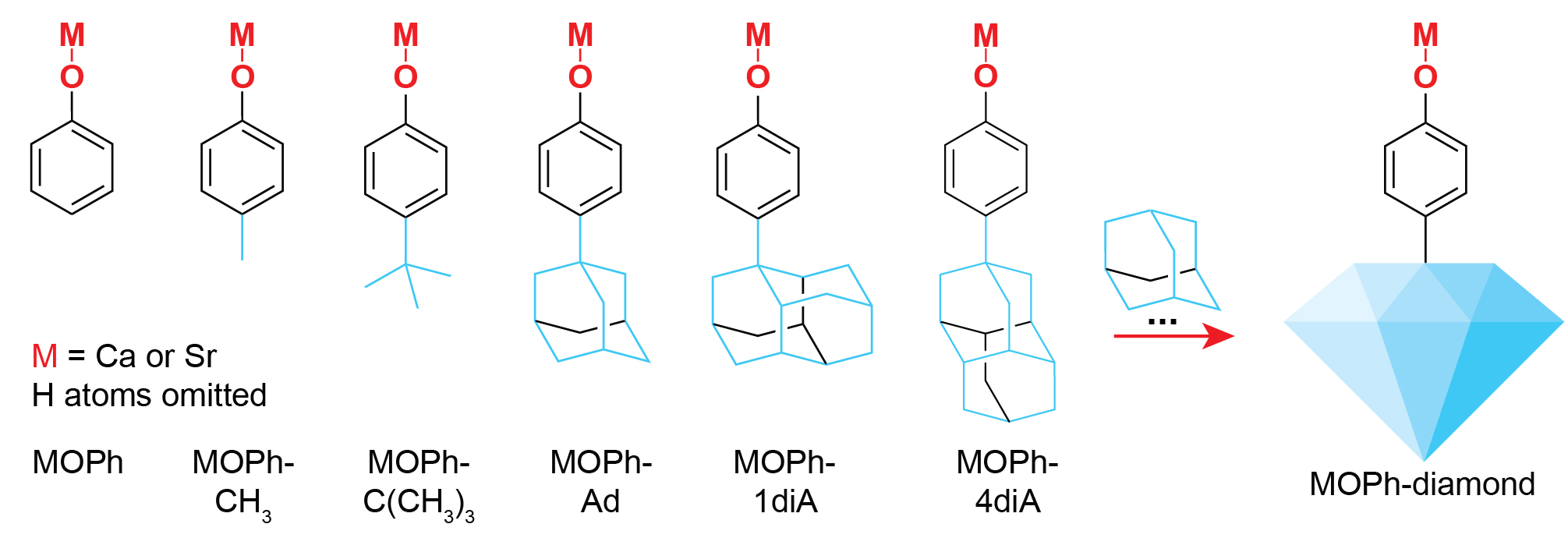}
    \caption{ Molecular structures of MOPh and all derivatives, showcasing a systematic growth in ligand size from -CH$_3$ to -diamantane 
    at the \textit{para} position. The addition of more adamantane units is indicative of progression towards a diamond structure. Metal M represents either Ca or Sr.
    }
   \label{scheme}
\end{scheme*}

To address these questions, we report a systematic, experimental and theoretical, bottom-up investigation of Ca/Sr-containing phenoxide molecules MOPh-X (M = Ca or Sr, Ph=phenyl, X=substituents) with increasing size of substituents. As depicted in Scheme \ref{scheme}, utilizing calcium and strontium phenoxides as modular OCCs, we substitute the \textit{para} position of phenyl ring with various
substituents, including alkyl groups -CH$_3$ and -C(CH$_3$)$_3$, and diamondoids adamantane, 1-diamantane, and 4-diamantane, with the surface of hydrogen-terminated bulk diamond investigated only theoretically. This allows us to investigate the transition from gas-phase to surface-bound OCC molecules. While MOPh-diamond was exclusively examined theoretically, all other five derivatives were produced in a cryogenic buffer gas cell and studied using dispersed laser-induced fluorescence (DLIF) spectroscopy. Despite significant structural variations, the transition energies of five derivatives remained within a narrow range of 10 \cm, suggesting consistent transition energies across diamond-based OCCs. The vibrational branching fractions are in the range of 0.86-0.95, similar to other phenoxides studied previously \cite{zhu2022functionalizing,lao2022sroph}, and show no trend with the size of the diamondoid. This, and the similarities of transition energies and molecular orbitals between isolated molecules and surface-bound MOPh OCCs underscore the promise of surface-bound OCCs as candidates for QIS applications.

Two distinct methods were used to produce all derivatives based on the melting points of precursors. For precursors with low melting points ($<100^{\circ}$C), including \textit{p}-cresol (HOPh-CH$_3$) and 4-\textit{tert}-butylphenol (HOPh-C(CH$_3$)$_3$), we utilized a gas-phase method \cite{zhu2022functionalizing,lao2022sroph} involving the reaction of metastable metal atoms with volatile ligands. These metal atoms were generated via laser ablation of calcium or strontium metal pellets within a cryogenic buffer gas cell. Conversely, for derivatives with non-volatile diamondoid-containing precursors, laser ablation of a mixture of metal hydrides (MH$_2$) and synthesized precursor ligands, using silver powder as a cohesive binder, was used  \cite{mitra2022pathway}.  These production approaches facilitate the creation of a diverse set of molecules with varying volatility and sizes. The metal-containing products were then brought to their vibrational ground states through neon buffer-gas collisions, and subsequently excited to higher electronic states using a pulsed dye laser. The emitted fluorescence was collected, dispersed via a monochromator, and recorded using an intensified charge-coupled device camera. Comprehensive experimental details and theoretical calculations of all molecules and surfaces are available in the Supplementary Information.

\begin{figure*}
    \centering
    \includegraphics[scale=0.45]{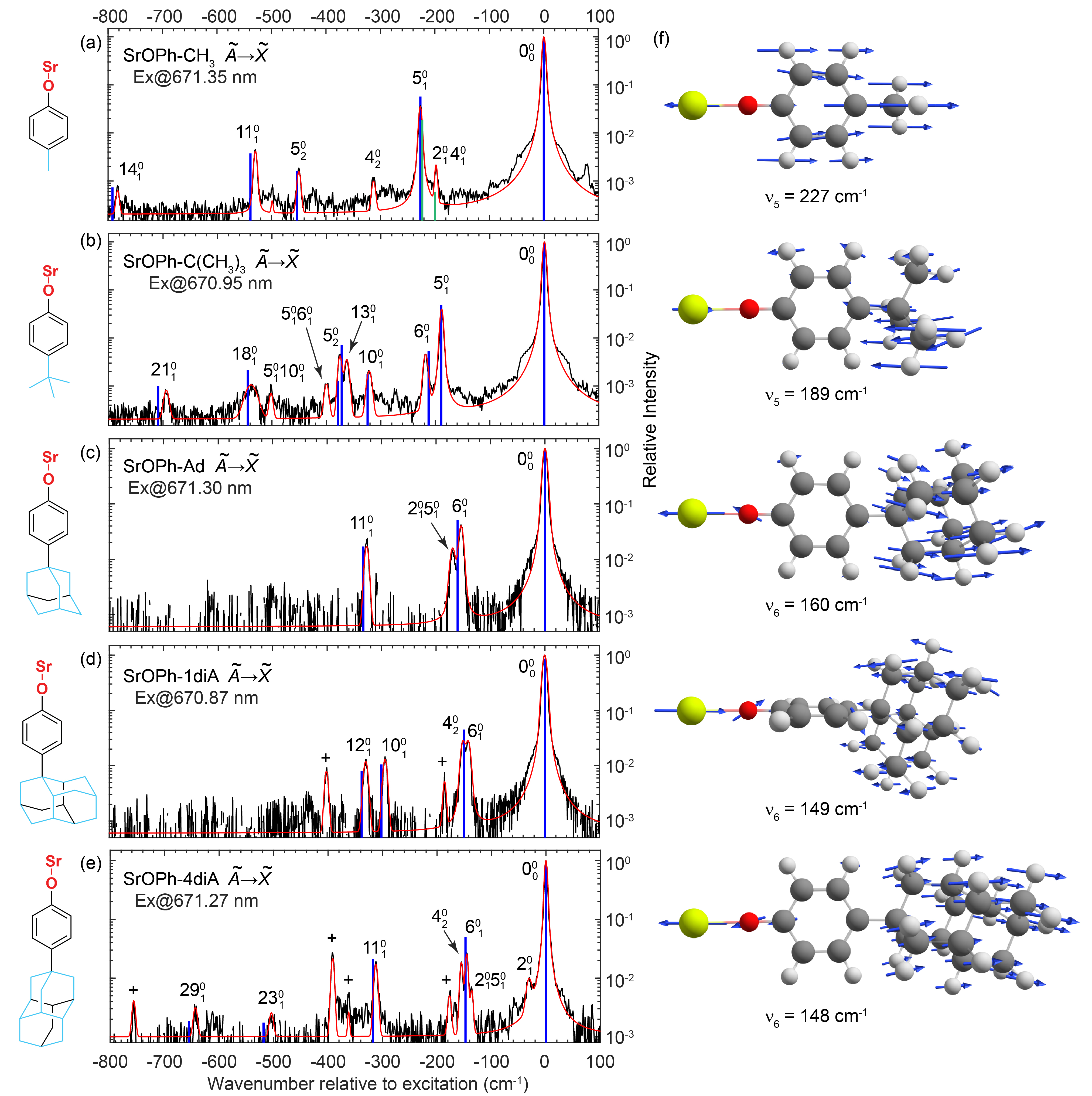}
    \caption{DLIF spectra of the $\widetilde A \rightarrow \widetilde X$ transitions of (a) SrOPh-CH$_3$, (b) SrOPh-C(CH$_3$)$_3$, (c) SrOPh-Ad, (d) SrOPh-1diA and (e) SrOPh-4diA. The experimental curves (black) are overlaid with the Voigt profile fits (red). The blue vertical lines indicate the calculated harmonic frequencies of the vibrational decays and the respective heights show the relative intensities from calculation. The two green sticks near -200 \cm in (a) represent the vibrational perturbation theory calculation of the Fermi resonance coupling between the fundamental mode $\nu_5$ and the combination mode $\nu_2+\nu_4$. ``+" labels denote strontium atomic lines produced from laser ablation of SrH$_2$. (f) The vibrational displacements and calculated harmonic frequencies of the lowest Sr-O stretching mode, the strongest off-diagonal vibrational decay.  
    } 
    \label{fig1:sroph-AX}
\end{figure*}

\section*{Results and discussion} 

\noindent{\textbf{DLIF spectra.}} The DLIF spectra of un-substituted CaOPh and SrOPh at cryogenic temperatures have been previously reported \cite{zhu2022functionalizing,lao2022sroph,zhu2024extending}. These spectra revealed two lowest electronically excited states \A $^2B_2$ and \B~$^2B_1$, which are proposed for laser cooling. The respective branching fractions for vibrationless decays were measured to be $\approx$ 0.90. In this study, we identified both states for all derivative molecules and recorded their respective DLIF spectra. For Sr-containing molecules, five spectra for the $\widetilde A \rightarrow \widetilde X$ transitions are shown in Fig. \ref{fig1:sroph-AX}.  Similar spectra for $\widetilde B \rightarrow \widetilde X$ and both transitions for Ca-containing molecules are presented in Figs.\ref{figs:Sr_BX_DLIF} and \ref{figs:Ca-DLIF}. All spectra are presented relative to their excitation energies, with peak intensities normalized to the peak at the origin. The relative peak intensities and frequency shifts are compared with the calculated harmonic frequencies (Tables \ref{tab:theo-freq-Ca}-\ref{tab:theo-freq-Sr}) and Franck-Condon factors (FCFs, Tables \ref{tab:theo-fcf-caophch3-tub}-\ref{tab:theo-fcf-sroph4diA}) for the assignment of vibrational modes. The peaks are fitted with Voigt functions to determine the peak areas for quantifying the intensity ratio of the observed vibrational decays.

Figure \ref{fig1:sroph-AX}a presents the dispersed spectrum for the $\widetilde A \rightarrow \widetilde X$ transition of SrOPh-CH$_3$ at an excitation wavelength of 671.35 nm. The peak at the origin, labeled as 0$_0^0$, represents the non-vibration-changing diagonal decay ($\widetilde A, \nu = n' \rightarrow \widetilde X, \nu = n''$). It is primarily from ground state $0' \rightarrow 0''$ decay, 
with minor contributions ($<1$\%) from higher vibrational levels $n'\rightarrow n'' (n>0)$ decays 
due to hot band excitations \cite{zhu2022functionalizing}. The strongest vibration-changing off-diagonal decay is observed at -227 \cm, corresponding to the theoretical harmonic frequency of the lowest Sr-O and ring stretching mode $\nu_5 = 227$ \cm, as shown in Fig. \ref{fig1:sroph-AX}(f). The observed intensity matches well with the calculated values depicted by blue sticks. In line with previous findings for CaOPh and SrOPh molecules, as well as their derivatives,  \cite{zhu2022functionalizing,lao2022sroph,zhu2024extending} this particular stretching mode, involving the metal-oxygen bond, is consistently the most prominent off-diagonal decay due to the predominant localization of molecular orbitals of both \X and \A states on the metal atoms \cite{dickerson2021optical,zhu2022functionalizing,lao2022sroph,zhu2024extending}. An unexpected red side peak at -198 \cm, not predicted under harmonic approximation, is accounted for by vibrational perturbation theory (VPT) calculation. \cite{pyVibPtboyer2021,boyer2022,VPT2-1} It is attributed to a combination mode $\nu_2 + \nu_4$ (194 \cm), resulting from the intensity borrowing through Fermi resonance coupling \cite{zhu2024extending} with the fundamental mode $\nu_5$. In addition, the overtone of $\nu_5$ is identified at -450 \cm. Two weaker Sr-O stretching modes, $\nu_{11}$=539 \cm and $\nu_{14}$ = 792 \cm (Table \ref{tab:theo-freq-Sr} and Fig.~\ref{figs:vib-srophch3}), are discerned at -530 \cm and -782 \cm, respectively. The two remaining  peaks, at -198 \cm, -313 \cm,
are attributed to the combination bands $\nu_2 + \nu_4$ (194 \cm) and overtone mode $2\nu_4$ (316 \cm), respectively, when comparing to the theoretical frequencies in Table \ref{tab:theo-freq-Sr}.

With a larger substituent, SrOPh-C(CH$_3$)$_3$ shows a similar decay pattern (Fig. \ref{fig1:sroph-AX}(b)). Besides the diagonal decay at the origin, the most prominent off-diagonal decay is observed at -189 \cm, which is attributed to the lowest-frequency Sr-O stretching mode $\nu_5 = 189$ \cm (Fig. \ref{fig1:sroph-AX}(f)). Other Sr-O stretching modes, such as $\nu_{6}$=212 \cm, $\nu_{10}$ = 325 \cm, $\nu_{13}$=372 \cm and $\nu_{18}$ = 544 \cm and $\nu_{21}$=709 \cm (Table \ref{tab:theo-freq-Sr} and Fig. \ref{figs:vib-srophch3}), are observed at shifts of -218 \cm, -321 \cm, -363 \cm, -538 \cm and -694 \cm, respectively. The presence of additional vibrational decays is likely due to the flexible structure of the t-butyl moiety. 

\begin{figure*}
    \centering
    \includegraphics[scale=0.65]{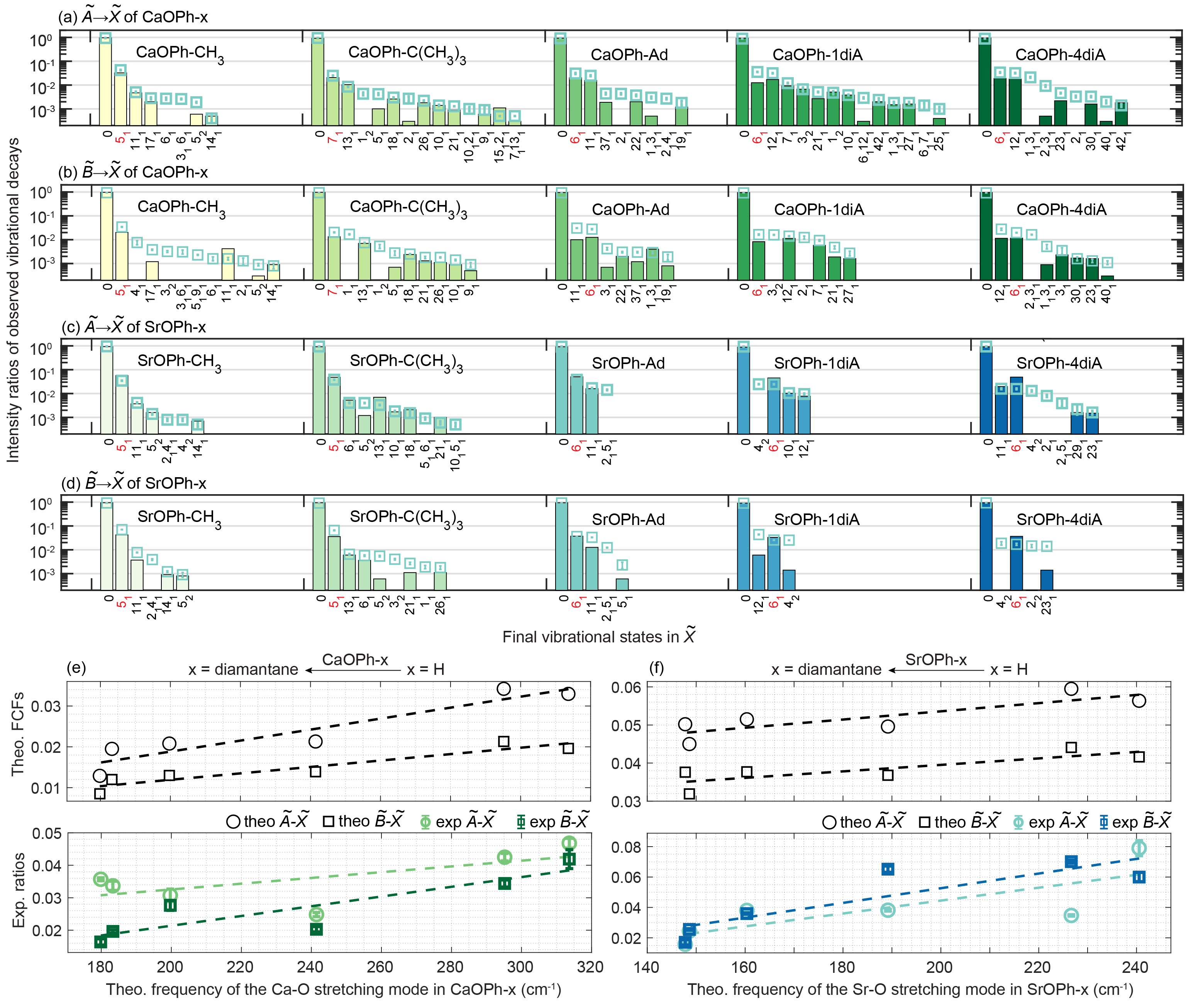}
    \caption{(a-d) Theoretical and experimental results for the intensity ratios of all observed vibrational decays of molecules studied in this work. Each molecule has both $\widetilde A \rightarrow \widetilde X$ and $\widetilde B \rightarrow \widetilde X$ transitions. The colored bars indicate the theoretical values while the empty squares show the experimental ratios extracted from the peak areas. Errors are statistical standard errors. All values are listed in Tables \ref{table:ca-vbr}-\ref{table:sr-vbr}. (e-f) Theoretical FCFs and experimental intensity ratios of the off-diagonal decay for the lowest-frequency metal-oxygen stretching modes (highlighted in red in a-d) in CaOPh-x and SrOPh-x molecules. The dash lines indicate general trends 
    suggesting that vibrational branching in these species does not degrade (and may even improve) as the size of the ligand increases.}
    \label{fig2:VBRs}
\end{figure*}

Figures \ref{fig1:sroph-AX}(c)-\ref{fig1:sroph-AX}(e) show the dispersed spectra for the $\widetilde A \rightarrow \widetilde X$ transitions in molecules with diamondoid substituents. Compared to the spectra of SrOPh-CH$_3$ and SrOPh-C(CH$_3$)$_3$, these spectra have a higher noise level ($\gtrsim 10^{-3}$) due to the lower production yield from the solid-phase method.
In Fig.~\ref{fig1:sroph-AX}(c), alongside the strong diagonal decay, three off-diagonal decays are observed. The doublet peaks at -160 \cm, assigned to the Sr-O stretching $\nu_6$ (Fig.~\ref{fig1:sroph-AX}(f)) and combination mode $\nu_2+\nu_5$, are also due to the Fermi resonance coupling \cite{zhu2024extending}. The fewer vibrational decays for the larger substituents are likely due to the rigid structure of adamantane. For the larger 1-diamantane substituent, the corresponding spectrum is slightly more complex (Fig.~\ref{fig1:sroph-AX}(d)). Besides the lowest-frequency stretching mode $\nu_6$ (149 \cm, Fig.~\ref{fig1:sroph-AX}(f)) at a frequency shift of -141 \cm, associated with the Fermi resonance coupling overtone mode $2\nu_4$ at -151 \cm, two other stretching modes containing the Sr-O bond are observed at -294 \cm and -329 \cm, which are respectively assigned to $\nu_{10}$ and $\nu_{12}$ (Table \ref{tab:theo-freq-Sr} and Fig.~\ref{figs:vib-srophAd}). Similar patterns are observed with the 4-diamantane substituent in Fig.~\ref{fig1:sroph-AX}(e). The frequency regions around -150 \cm, representing the vibrational decay of the lowest-frequency stretching mode, show multiple peaks. The middle strong peak at -146 \cm is from the stretching mode $\nu_6$ (Fig.~\ref{fig1:sroph-AX}(f)), while the other two peaks at -136 \cm and -155 \cm arise from Fermi resonance coupling from modes $\nu_2+\nu_5$ and $2\nu_4$, respectively. The peaks labeled with ``+" signs represent the strontium atomic lines at 679 nm ($5s6s~{}^3\mathrm{S}_1 \rightarrow 5s5p~{}^3\mathrm{P}_0^o)$, 687 nm ($5s6s~{}^3\mathrm{S}_1 \rightarrow 5s5p~{}^3\mathrm{P}_1^o)$, 689 nm ($5s5p~{}^3\mathrm{P}_1^o \rightarrow 5s^2~{}^1\mathrm{S}_0)$ and 707 nm ($5s6s~{}^3\mathrm{S}_1 \rightarrow 5s5p~{}^3\mathrm{P}_2^o)$, produced from the ablation of SrH$_2$ in the pressed target.

The dispersed spectra of $\widetilde B \rightarrow \widetilde X$ transitions of Sr-containing molecules in Fig.~\ref{figs:Sr_BX_DLIF} and those of Ca-containing molecules in Fig.~\ref{figs:Ca-DLIF} have their peak assignments detailed in the Supporting Information. Tables \ref{table:modes-freq}-\ref{table:sr-vbr} summarize the comprehensive vibrational frequencies and intensity fractions for all resolved vibrational decays. The displacements of all resolved fundamental vibrational modes are presented in Figs. \ref{figs:vib-srophch3}-\ref{figs:vib-caophdiA}.

\begin{figure*}
    \centering
    \includegraphics[scale=0.5]{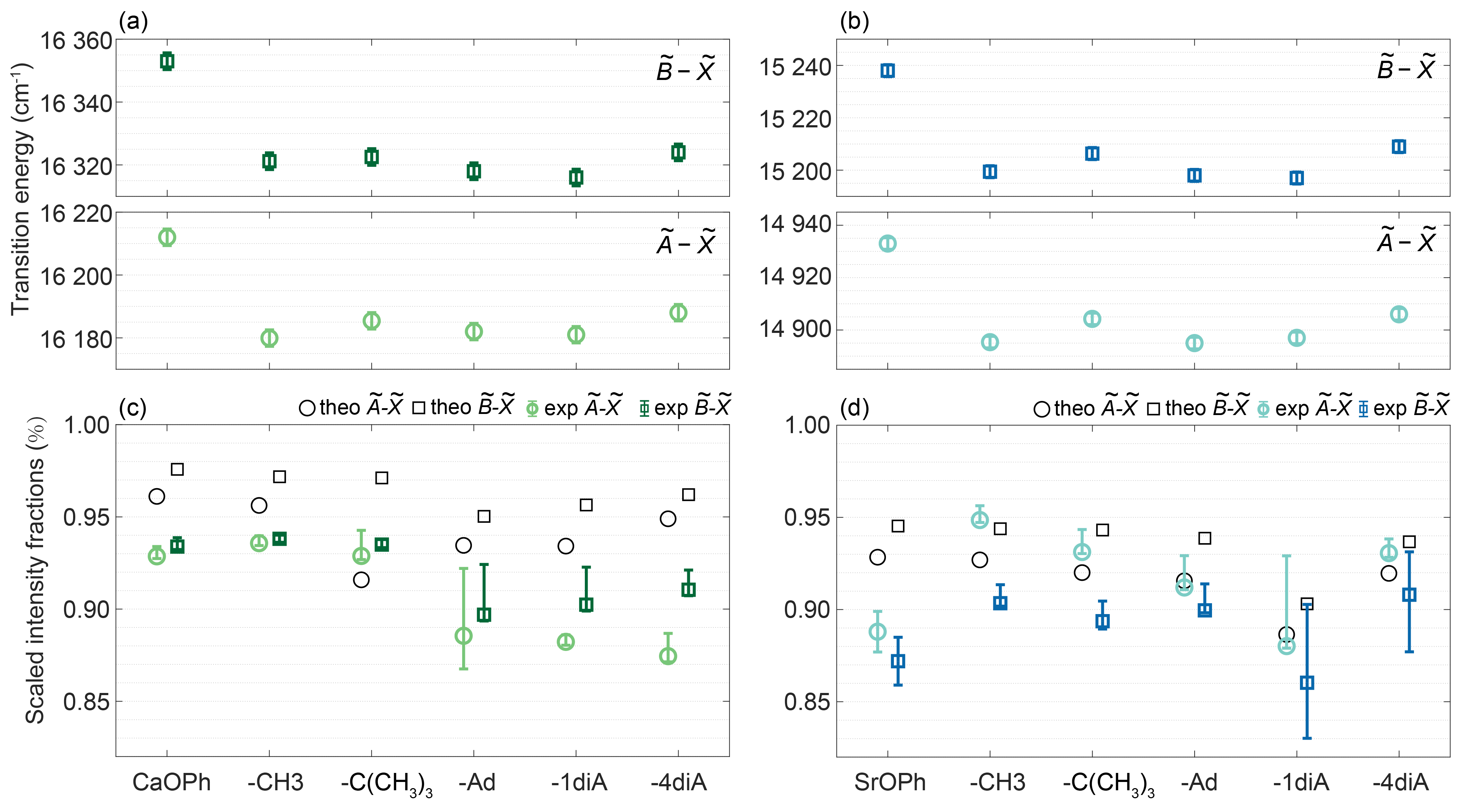}
    \caption{(a-b) Measured excitation energies of $\widetilde A \rightarrow \widetilde X$ and $\widetilde B \rightarrow \widetilde X$ transitions for all molecules studied in this work. 
    (c-d) Scaled experimental intensity fractions of the diagonal transitions for all molecules. Theoretical VBFs 
    are given for comparison.  All values are listed in Table \ref{tab:diagonal-VBRs} while those of CaOPh and SrOPh are taken from Refs.~\cite{zhu2022functionalizing,lao2022sroph,zhu2024extending}.
    } 
    \label{fig3:ex&vbr}
\end{figure*}

\vspace{0.5cm}
\noindent{\textbf{Intensity fractions of all vibrational decays.}}
Figures.~\ref{fig2:VBRs}a-d illustrates the intensity ratios of all vibrational decays. Generally, the experimental results align well with theoretical predictions, with some exceptions for overtone or combination modes related to Fermi resonance coupling, which eludes the harmonic approximation calculations. Most molecules exhibit only two or three significant vibrational decays with intensities above 10$^{-2}$, and major decays have intensity ratios between $10^{-3}$ and $10^{-2}$, demonstrating little variation despite the wide range of substituent size. According to the calculation results in Tables \ref{table:ca-vbr}-\ref{table:sr-vbr}, the sum of the vibrational branching fractions (VBFs) for the undetected vibrational decays with low VBF is approximately $0.3-5\%$, suggesting that OCCs could scatter about $20-300$ photons if all the measured vibrational transitions are addressed during repumping. 
Molecules with the -C(CH$_3)_3$ substituent exhibit a greater number of vibrational decays, likely due to their flexible structures.

Across all molecules, the most or second-most significant off-diagonal leakage is the lowest-frequency metal-oxygen stretching mode (highlighted in red in Figs.~\ref{fig2:VBRs}a-d). 
As the substituents become heavier, there is a consistent decrease in vibrational frequencies for this main leakage mode (Figs. \ref{fig1:sroph-AX}f). Furthermore, theoretical calculations show a possible trend for the FCFs of this main leakage to decrease with frequency, while the experimental intensity ratios show more scatter but are also consistent with this or no systematic trend, as shown in Figs.~\ref{fig2:VBRs}e-f.
When the substituent grows towards a bulk diamond structure, the lowest-frequency stretching modes in MOPh-diamond are calculated to have smaller vibrational frequencies (Fig. \ref{figs:vib-diamond}), and similar or smaller off-diagonal FCFs are expected.

\begin{figure}
    \centering
    \includegraphics[scale=0.45]{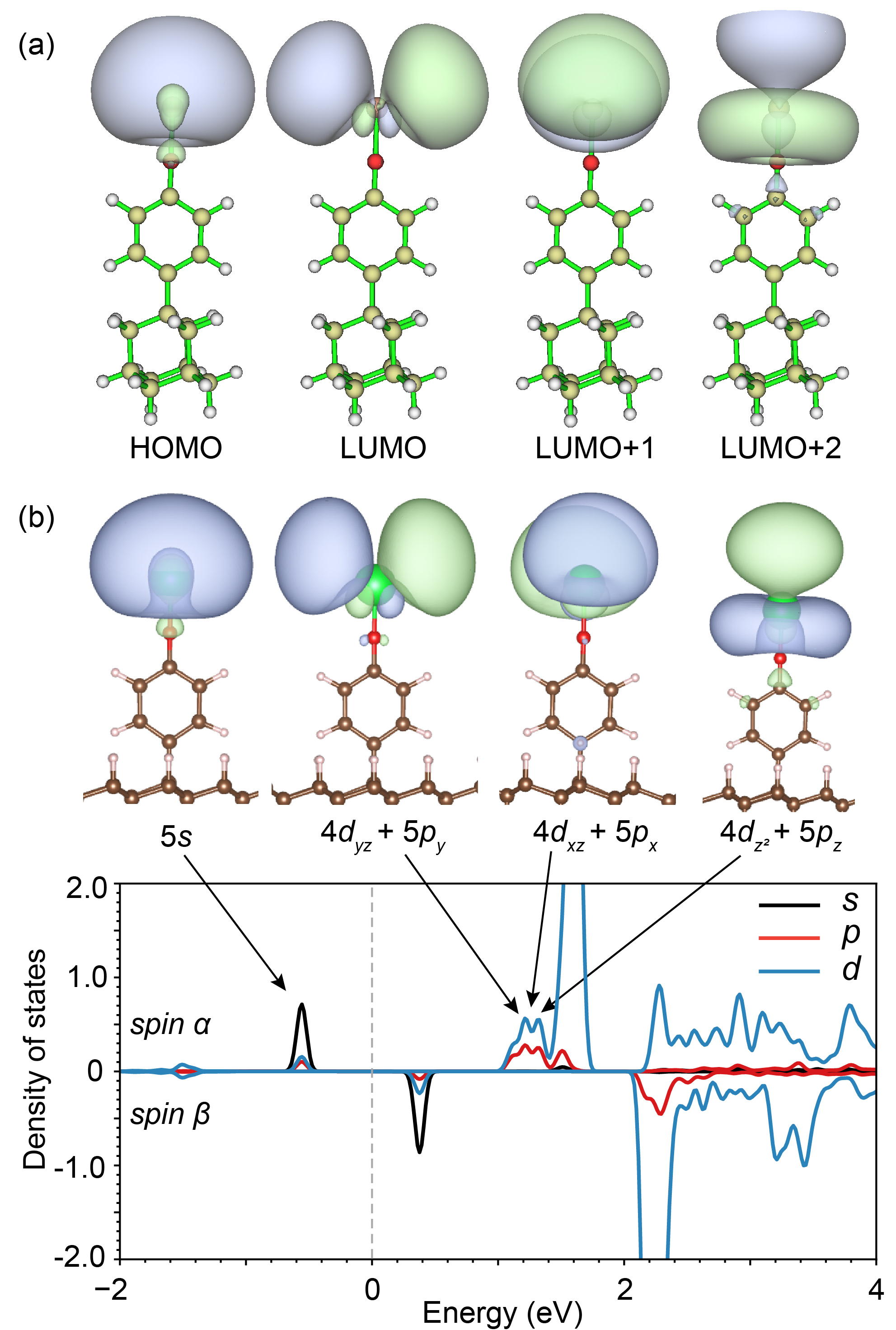}
    \caption{
    (a) Molecular orbitals of HOMO and LUMOs of SrOPh-Ad. (b) Projected density of states (PDOS) and molecular orbitals of the low-energy states of SrOPh-diamond. The PDOS shows the contributions from Sr \textit{s, p, d} orbitals and the OCC orbitals indicate the orbital hybridization and the electronic structure in SrOPh-diamond. The full PDOSs of both CaOPh-diamond and SrOPh-diamond are presented in Fig. \ref{figs:ca-sr-pdos}.
    } 
    \label{fig:PDOS}
\end{figure}

\vspace{0.5cm}
\noindent{\textbf{Transition energies and diagonal VBFs.}} 
To further investigate the influence of substituent size and complexity on OCC, we analyzed the transition energies and intensity ratios of diagonal vibrational decays. 
Figs.~\ref{fig3:ex&vbr}(a)-(b) show that the transition energies of CaOPh and SrOPh are approximately 40 \cm higher than those of their derivatives, due to the electron-donating nature of hydrocarbon ligands, in contrast to the strong electron-withdrawing characteristics of groups like -F or -CF$_3$ \cite{zhu2022functionalizing,lao2022sroph}.
The minimal variation in transition energies, remaining within 10 \cm for all derivatives, underscores a consistent electronic structure across different substituents.
Projecting this pattern forward, we anticipate that the transition energies for substituents of larger diamondoids and bulk diamonds will remain close.
This hypothesis is supported by our theoretical embedding calculations. 
Using a quantum-in-quantum embedded electronic structure technique devised by some of the authors \cite{lavroff2024aperiodic}
, vertical excitation energies for the CaOPh moiety on the diamond (111) surface were calculated at the NEVPT2(9,10)\cite{Angeli2007} level. 
The results show vertical excitations from the ground state, $\widetilde X$, to the quasi-degenerate, \textit{p}-type first and second excited states, $\widetilde A$ and $\widetilde B$ (Fig.~\ref{fig:PDOS}(a)), with the energies of 1.93 eV and 1.95 eV, respectively.
These findings from embedding calculations are close to the theoretical and experimental values for diamondoids, reinforcing the expectation of consistent transition energies in larger OCC structures.
This constancy of the transition energy in the diamonoid series makes it easy to identify the excitation transition in MOPh-diamond complexes.

Figs.~\ref{fig3:ex&vbr}c-d illustrate the scaled
intensity ratios of the diagonal vibrational decay (see SI) for Ca- and Sr-containing molecules, respectively.
For Ca derivatives, the experimental ratios were observed to range $\approx 0.88-0.94$, consistently lower than the theoretical predictions.
For Sr-containing molecules, the experimental ratio range is slightly broader, spanning from 0.86 to 0.95, and with better agreement with theory.
These diagonal ratios are comparable to other calcium and strontium phenoxides and derivatives, \cite{zhu2022functionalizing,mitra2022pathway,lao2022sroph}
suggesting that larger diamondoid substituents have a negligible impact on the diagonal decays. This is because the highest occupied molecular orbital (HOMO) and lowest unoccupied molecular orbital (LUMO) are primarily localized on the metal atoms, irrespective of substituent size and complexity, as shown for SrOPh-Ad molecule in Fig.~\ref{fig:PDOS}a. 

With an extension to surfaces of hydrogen-terminated diamond, theoretical calculations of the projected density of states (PDOS) for SrOPh-diamond indicate that the molecular orbitals of the occupied and unoccupied higher states closely resemble those in isolated molecules (compare Figs.~\ref{fig:PDOS}(a) and (b)), with the electron density primarily localized on the metal atom. The orbital mixtures are identical to those in isolated molecules, suggesting that SrOPh-diamond will show similar optical cycle closure to the gas-phase species. Compared to SrO-diamond, a bare OCC on a diamond, \cite{guo2021surface}, the phenyl ring in MOPh-diamond would separate OCCs from the diamond surfaces and possess good electron-withdrawing properties.

\noindent{\textbf{Discussion and prospects.}} The DLIF spectra of the $\widetilde A \rightarrow \widetilde X$ transitions in strontium phenol-adamantanes
reveal that the diagonal vibrational branching fractions exceed 90\%, with no discernable trend as the diamondoid substituents are increased in size (atom number) over roughly an order of magnitude.  Of particular interest is the fact that this dynamic range in substituent size also more than triples the number of vibrational modes the molecule has, a scenario that recent analysis of the role of vibrational coupling suggests could have been problematic for optical cycle closure \cite{Zhang2023IntensityBorrowing,zhu2024extending}.  However, since the main leakage channels will involve significant stretch character for the metal atom, the addition of modes with little overlap with that displacement does not lead to additional suppression of the vibrationless decays.

Along with our theoretical calculations, this finding suggests that very large or even diamond-surface-bound OCCs will be capable of scattering multiple photons using only a few pumping lasers. Notably, the vibrational decays observed correspond predominantly to a limited number of stretching and bending modes. This suggests that achieving vibrational cooling for surface-bound OCCs could be efficiently managed with only a few repumping lasers. Furthermore, the constrained rotational degrees of freedom may allow for rotational cooling through pumping the P-branch transitions $K\rightarrow K-1$ of the surface-bound OCCs, where $K$ is the projection quantum number of the rotational angular momentum onto the axis perpendicular to the diamond surface. 

Aside from their optical cycling properties, the multiple quantum degrees of freedom of surface-bound OCCs can be another advantage for QIP. For example, in the $\Tilde{X}$ state, due to the spin-rotational and spin-spin interactions, one can define qubits with spins ($S$ or $I$) and couple them to the rotational states $K$ for quantum error correction \cite{jain2023ae}. Hyperfine qubits can be furnished by a spinful metal isotope for the OCC, such as $^{43}$Ca and $^{89}$Sr, and readout would be performed by state-selective LIF. These quantum degrees of freedom of the surface-OCCs open new avenues for quantum manipulation and storage, potentially including the development of large-scale quantum computers that may surpass current capabilities in terms of qubit numbers by several orders of magnitude.

Moreover, MOPh OCC molecules can be densely packed to provide an exceptional platform to investigate the collective dynamics of an ensemble of emitters, including phenomena such as superradiant and subradiant effects \cite{reitz2022cooperative,trebbia2022tailoring}. Surface-bound OCCs can be synthesized through chemical functionalization on hydrogen-terminated diamond surfaces, as shown in Scheme \ref{scheme2} (and SI section \ref{sec:surface occ}). The hydrogen atoms terminating the diamond surface can be readily substituted with oxygen, nitrogen and sulfur atoms by various chemical treatments \cite{2008diamondfunction,2019diamondfunction}. This versatility allows for the functionalization of the diamond surface with a wide range of bond types and ligand molecules, enabling tailored modifications to suit specific applications.

\section*{Conclusion} 
Calculations and measurements have shown that the diagonal VBFs of molecules can remain high ($>0.85$) even as the vibrational mode density increases over a wide dynamic range.  The probability of diagonal vibrational decays does not systematically decrease with increasing molecular size or increasing number of potential vibrational transition channels of the molecule. These results demonstrate that optical cycling is not necessarily compromised by simply adding vibrational modes, and highlight the fact that the identities of the modes can be more important than their sheer number in determining optical cycle closure.  These observations pave the way to significantly larger species capable of being laser cooled and manipulated in single quantum states than the current state of the art, including the possibility of surface-bound qubits.

\section*{Acknowledgements}
This work was supported by the NSF Center for Chemical Innovation Phase I (grant no. CHE-2221453), AFOSR (grant no. FA9550-20-1-0323), the NSF (grant no. OMA-2016245, PHY-2207985 and DGE-2034835). This research is funded in part by the Gordon and Betty Moore Foundation (DOI: 10.37807/ GBMF11566).  Computational resources were provided by XSEDE and UCLA IDRE shared cluster hoffman2. The authors acknowledge computational resources from the National Energy Research Scientific Computing Center (NERSC), a U.S. Department of Energy Office of Science User Facility.

\newpage
\bibliographystyle{apsrev4-1_no_Arxiv}
\bibliography{ref}

\renewcommand*{\thefigure}{S\arabic{figure}}
\setcounter{figure}{0} 

\begin{figure*}
\centering
    \includegraphics[scale=0.45]{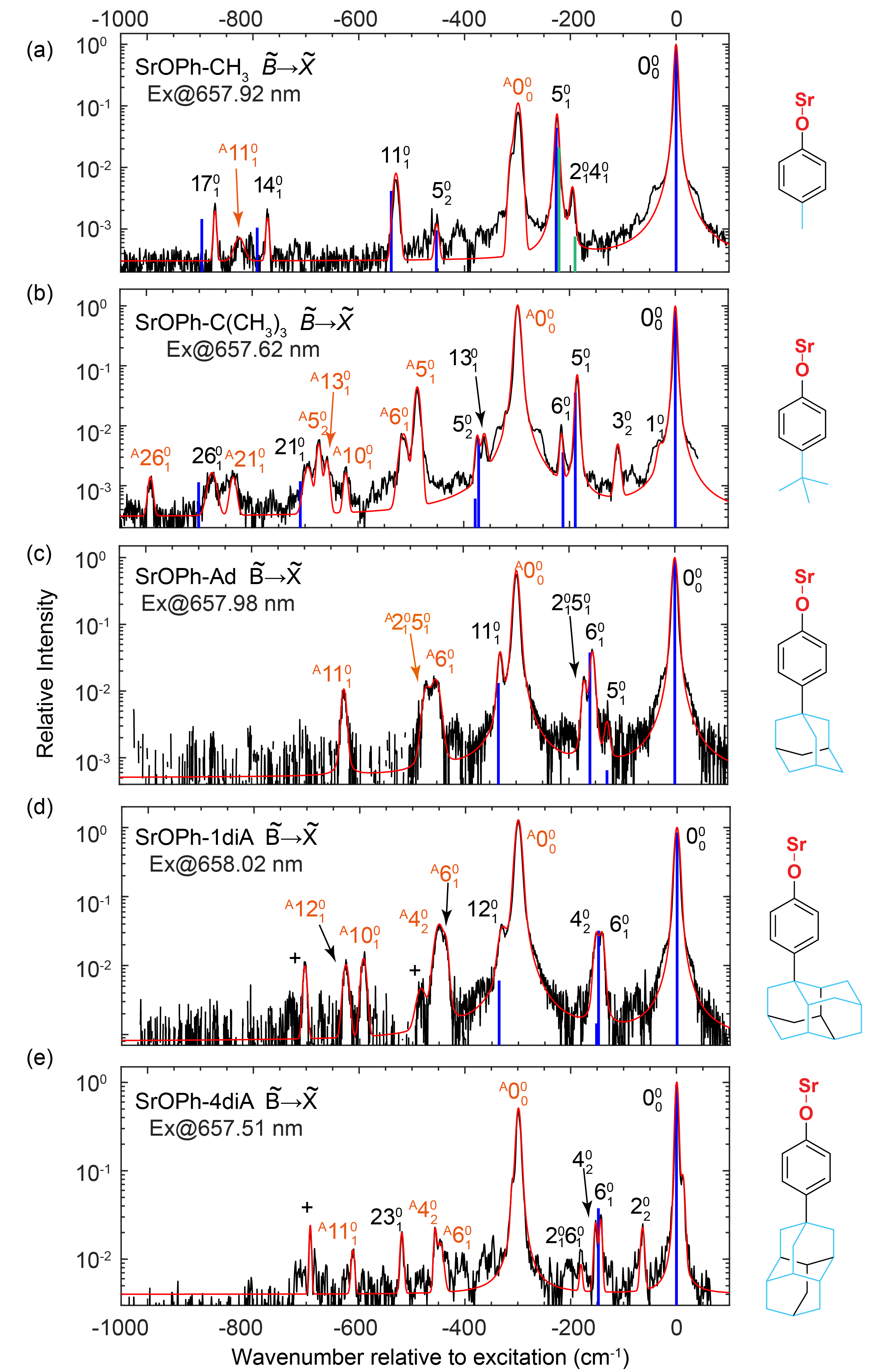}
    \caption{Dispersed spectra of the $\widetilde B \rightarrow \widetilde X$ transitions of (a) SrOPh-CH$_3$, (b) SrOPh-C(CH$_3$)$_3$, (c) SrOPh-Ad, (d) SrOPh-1diA and (e) SrOPh-4diA. The black and red curves show the experimental spectra and the Voigt profile fits, respectively. The blue sticks indicate the theoretical predictions of the intensity ratios and harmonic frequencies of the vibrational decays. The lines labeled with the $+$ signs correspond to the strontium atomic lines.}
    \label{figs:Sr_BX_DLIF}
\end{figure*}

\begin{figure*}
    \centering
    \includegraphics[scale=0.45]{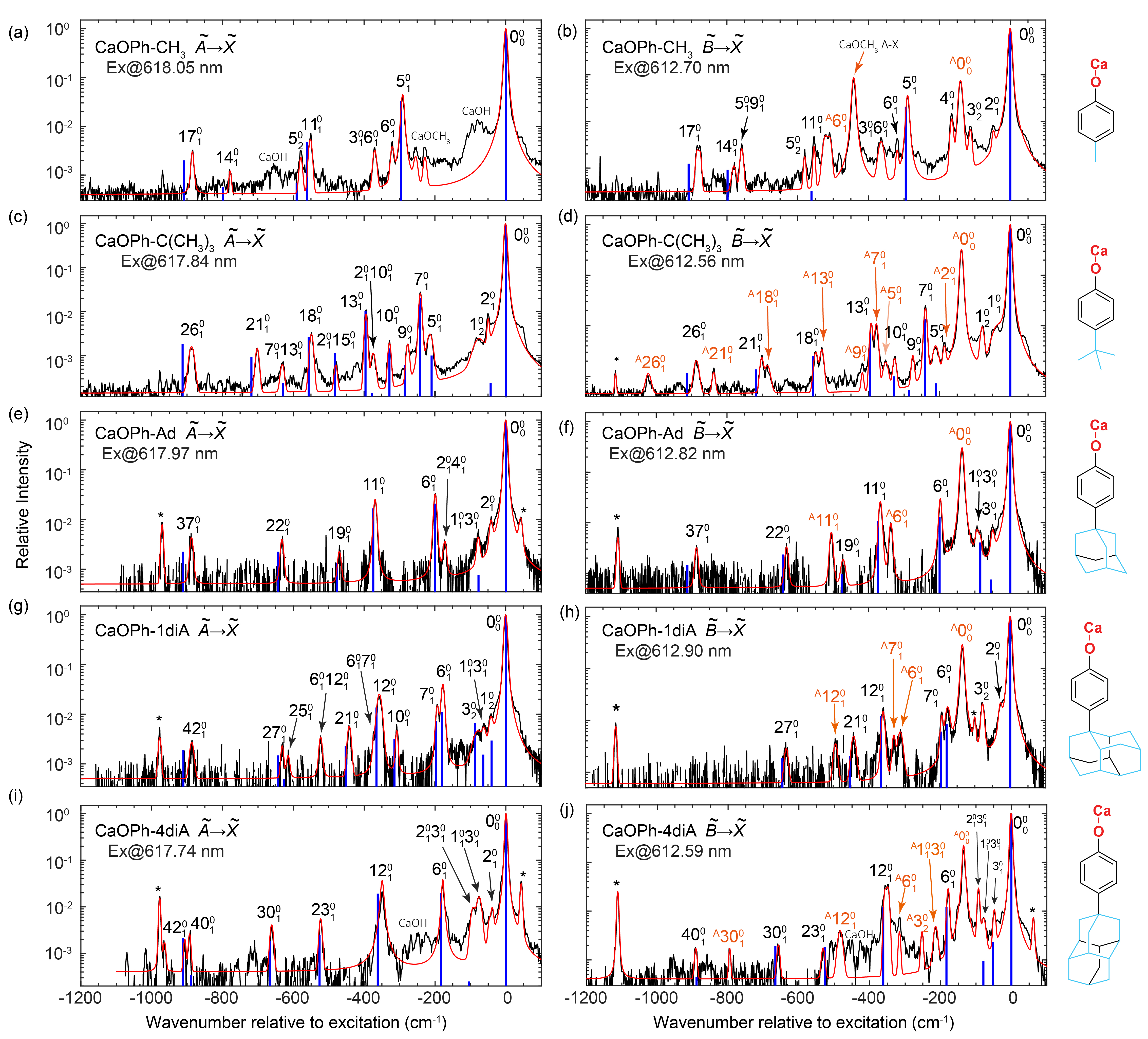}
    \caption{Dispersed spectra of the $\widetilde A/\widetilde B \rightarrow \widetilde X$ transitions of (a-b) CaOPh-CH$_3$, (c-d) CaOPh-C(CH$_3$)$_3$, (e-f) CaOPh-Ad, (g-h) CaOPh-1diA and (i-j) CaOPh-4diA. The black and red curves show the experimental spectra and the Voigt profile fits, respectively. The blue sticks indicate the theoretical predictions of the intensity ratios and harmonic frequencies of the vibrational decays. 
    $*$ indicates the calcium atomic lines induced by the laser ablation of the calcium/CaH$_2$ targets.
    }
    \label{figs:Ca-DLIF}
\end{figure*}

\section*{Supplementary information}

\subsection{Spectroscopy measurement}
The molecules involved in the present work were produced by the reaction of ligand precursors, including 4-methylphenol, 4-tert-Butylphenol, 4-(1-Adamantyl)phenol, 4-(1-diamantyl)-phenol and 4-(4-diamantyl)phenol, with metal atoms (Ca/Sr) or hydrides (CaH$_2$/SrH$_2$) in the cryogenic buffer gas cell. The first three precursors were purchased from Sigma Aldrich while the rest two ligands were synthesized according to section \ref{sec:synthesis}. Two different methods are used to load the ligand precursors into the cryogenic cell. One method is gas phase hot vapor loading \cite{zhu2022functionalizing,lao2022sroph,zhu2024extending}. The volatile organic ligands, including 4-methylphenol and 4-tert-Butylphenol, in a separated reservoir were heated to the melting point. The hot ligand vapors were flowing into the cell directly or carried through helium gas at a flow rate of approximately 0.2 sccm. They could react with meta-stable metal atoms produced by ablating Sr or Ca metal pellets to form the corresponding products. An Minilite II 1064nm pulsed Nd:YAG laser (repetition rate 10 Hz) at pulse energy $\approx 6 $ mJ was used for ablation. The gas line was heated to $>120$ $^{\circ}$C to prevent the vapor from freezing in the line. The number density of the ligand precursors in the cell is $\approx 10^{12}-10^{13}$ cm$^{-3}$. The reaction products were cooled by collision with neon buffer gas of density $\approx 10^{15}-10^{16}$ cm$^{-3}$. The other solid method, utilizing a direct ablation of mixture targets,\cite{mitra2022pathway} is used for the low volatile precursors such as 4-(1-Adamantyl)-phenol, 4-(1-diamantyl)-phenol and 4-(4-diamantyl)-phenol. The ligand precursors with the dihydride of the metal (SrH$_2$ or CaH$_2$, powder) and silver powder (serves as the binder) in a mass ratio of \(m_{\text{ligand}}:m_{\text{MH}_2}:m_{\text{Ag}} \approx 1:1:3\) and then pressed the mixture into an ablation target. In practice, as the ablation of the composite target in the cell can easily vaporize these ligand precursors with high melting points (estimated to be over $150^{\circ}$C for Phenol-Ad and Phenol-diA), we found that this method can effectively integrate the ligands into the reaction. A higher ablation energy ($15-20 $ mJ) was essential to achieve a good signal to noise ratio.

About $0.7-2$ ms delay after the ablation, the cooled molecules were then optically pumped to the excited states by a tunable, pulsed dye laser (10 Hz, LiopStar-E dye laser, linewidth 0.04 cm$^{-1}$ at 620 nm). Here, the delay time was set to maximize the fluorescence signal strength, which depends mostly on the experiment configuration, such as buffer gas flow rate (typical setting is $\approx20$ sccm) and the molecular species. 
The method of searching for these excited states can be found in the SI of previous work\cite{lao2022sroph}. The fluorescence from the excited states was collected via an imaging system 
into a model 2035 McPherson monochromator equipped with a 1200 lines/mm grating. An Andor iStar 334T intensified charge-coupled device camera (ICCD) was used to record the dispersed spectra, with its gate time set as $\approx 20-170$ ns after the laser pulse for capturing the fluorescence signal right after the excitation (pulse duration $\approx20$ ns) while precluding signals of the PDL and ambient light. The entrance slit width was set at $\approx$ 0.10mm, resulting in a resolution of $\approx$ 10 cm$^{-1}$ in the spectra. 

\subsection{Synthesis of ligand precursors}
\label{sec:synthesis}
All materials were of analytical grade and purchased and used as received from Fisher Scientific, Acros Organics, Sigma Aldrich, TCI Chemicals AK Scientific, and Oakwood Chemicals. Diamantane was obtained as a gift from Chevron. Unless otherwise noted, all reactions were carried out in flame-dried glassware equipped with a stir bar and under an atmosphere of argon. NMR spectra were recorded on a DRX 400 Bruker spectrometer at 400 MHz ($^1$H) and 101 MHz ($^{13}$C). NMR spectra (Figs. \ref{figs:1diA-HNMR}-\ref{figs:4diA-CNMR}) were referenced to residual solvent peak in deuterated chloroform (Cambridge Isotope Laboratories). IR spectra (Figs. \ref{figs:1diA-IR}-\ref{figs:4diA-IR}) were measured on a Perkin-Elmer Spectrum Two FT-IR Spectrometer. High-resolution mass spectrum data was measured on a DART spectrometer. Melting point values were recorded using a Melt-Point II apparatus. Column chromatography was performed using silica gel (Millipore, 60 $\mathring{A}$, $0.0063-200$ mm, $70-230$ mesh) as stationary phase. Merck silica gel (60 F254) or neutral alumina (150 F254) sheets were used for thin layer chromatography (TLC).  

\subsubsection*{\normalfont{4-(1-diamantyl)-phenol (Scheme \ref{syn-2yl})}}

\renewcommand*{\thescheme}{S\arabic{scheme}}
\setcounter{scheme}{0} 
\begin{scheme*}
    \centering
    \includegraphics[scale=0.6]{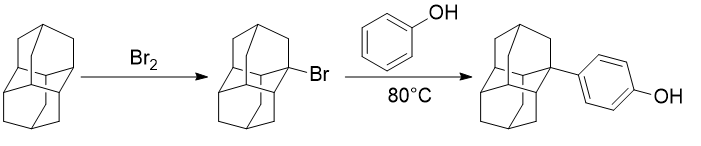}
    \caption{Synthesis procedure of 4-(1-diamantyl)-phenol.
    }
   \label{syn-2yl}
\end{scheme*}

\begin{scheme*}
    \centering
    \includegraphics[scale=0.6]{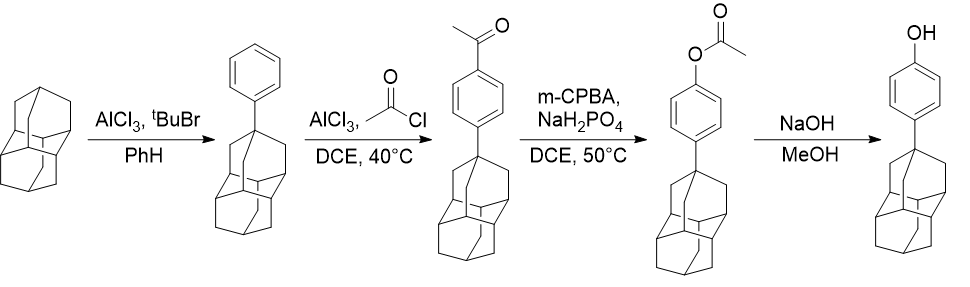}
    \caption{Synthesis procedure of 4-(4-diamantyl)-phenol.
    }
   \label{syn-4yl}
\end{scheme*}

\noindent\textit{1-bromodiamantane}: 1-bromodiamantane was prepared using similar conditions published by Gund \textit{et al}.\cite{gund1974diamantane} After working up the reaction, the solvent was removed under vacuum, and the resulting residue was concentrated thrice from hexanes to afford a yellow solid. The solid was dissolved in 9/1 hexane/dichloromethane (DCM) and was passed through a silica plug to afford of white crystalline solid. Yield: 1.34g (95\%). Characterization data matches that of literature.\cite{gund1974diamantane} $R_f=0.77$ (silica, hexanes) $^1$H NMR (400 MHz, CDCl$_3$) $\delta$~2.47-2.42 (m, 4H), 2.13 (m, 2H), 2.06 (m, 2H), 1.96-1.91 (m, 1H), 1.82-1.77 (m, 1H), 1.77-1.73 (m, 2H), 1.72 (m, 4H), 1.68 (m, 1H), 1.61 (m, 1H), 1.58 (m, 1H). $^13$C NMR (101 MHz, CDCl$_3$) $\delta$~79.47, 51.69, 46.21, 41.58, 38.67, 37.31, 36.55, 34.86, 31.52, 25.25.

\smallskip
\noindent\textit{\textbf{4-(1-diamantyl)-phenol}}: 1-bromodiamantane (100 mg, 0.37 mmol) and phenol (282mg, 3 mmol) were combined and purged thrice with argon. The solids were heated at 80 $^{\circ}$C for 40 min, at which time the reaction dried up to afford a pink solid at the bottom of the flask. The solid was suspended in hot water, filtered and washed thrice with hot water. The solid was then dissolved in ethyl acetate and dried over NaSO$_4$. The solvent was removed under vacuum and the crude mixture was purified on silica (10/1 hexane/ethyl acetate) to afford 60 mg of white crystalline solid. Yield: 58\%. m.p. 211-213$ ^\circ C$. $R_f=0.17$ (silica, 10/1 hexanes/ethyl acetate). $^1$H NMR (400 MHz, CDCl$_3$) $\delta$~7.23-7.18 (m, 2H), 6.82-6.76 (m, 2H), 4.59 (s, 1H), 2.32-2.26 (m, 2H), 2.00-1.95 (m, 2H), 1.89-1.84 (m, 2H), 1.84-1.81 (m, 1H), 1.81-1.75 (m, 5H), 1.69-1.63 (m, 3H), 1.56-1.50 (m, 2H), 1.45-1.37 (m, 2H). $^13$C NMR (101 MHz, CDCl$_3$) $\delta$~152.71, 141.82, 126.55, 115.10, 49.72, 41.46, 38.50, 38.47, 38.33, 38.13, 37.17, 33.55, 28.12, 26.06. IR (neat) $\nu_{max}$= 3283, 2907, 2886, 2849, 1597, 1511, 1439, 1248, 808 cm$^{-1}$. HRSM (DART) calcd. for [C$_{20}$H$_{22}$O$+$H]$^+$ 279.17434, found 279.17439.

\subsubsection*{\normalfont{4-(4-diamantyl)-phenol (Scheme \ref{syn-4yl})}}

\noindent\textit{4-phenyldiamantane}: Ground diamantane (1.88 g, 10 mmol) and aluminum chloride (100 mg, 0.08 mmol) were suspended in 50 mL benzene, and tert-butyl bromide (0.66 mL, 5.5 mmol) was added slowly to the mixture. Gasses expelled were quenched with sat. NaHCO$_3$. After stirring for 30 min, another portion of tert-butyl bromide (0.66 mL) was added and the reaction was allowed to stir 30 min more before diluting with 50 mL diethyl ether. The solution was washed with 0.5 M HCl, water, brine and dried over Na$_2$SO$_4$. The filtrate was concentrated under vacuum, and crude mixture was purified on silica (hexane) to afford crystalline product. Yield: 1.3g (50\%). m.p. 145-147$ ^\circ C$. $R_f=0.68$ (silica, hexanes). $^1$H NMR (400 MHz, CDCl$_3$) $\delta$~7.45-7.38 (m, 2H), 7.37-7.31 (m, 2H), 7.23-7.17 (m, 1H), 1.99-1.93 (m, 3H), 1.93-1.89 (m, 6H), 1.89-1.83 (m, 1H), 1.83-1.77 (m, 9H). $^13$C NMR (101 MHz, CDCl$_3$) $\delta$~150.71, 128.05, 125.47, 125.08, 43.97, 38.27, 37.78, 36.74, 34.26, 25.76. IR (neat) $\nu_{max}$= 2906, 2882, 2848, 1493, 1459, 1442, 756, 696 cm$^{-1}$.

\smallskip %
\noindent\textit{4-(4-diamantyl)acetophenone}: Acetyl chloride (72 $\mu$L, 1 mmol) was added slowly to a stirring suspension of aluminum chloride (100 mg, 0.75 mmol)  in 2 mL dichloroethane at room temperature. 4-phenyldiamantane (132 mg, 0.5 mmol) in 0.5 mL chloroform was added dropwise to the resulting yellow solution and then the reaction was heated at 40$^{\circ}$C for 10 min. The reaction was then cooled down to room temperature and added into a stirring slurry of conc. HCl (1 ml) and ice (20 mL). The mixture was diluted with 20 mL chloroform, and the layers were separated. The organic layer was washed with sat. NaCO$_3$, water, brine, and dried over Na$_2$SO$_4$. Solvent was removed under vacuum and the crude material was purified on silica (95/5 hexane/ethyl acetate) to afford product. Yield: 100 mg (66\%). m.p. 195-203$ ^\circ C$. $R_f=0.35$ (silica, 10/1 hexanes/ethyl acetate). $^1$H NMR (400 MHz, CDCl$_3$) $\delta$~7.94-7.88 (m, 2H), 7.50-7.44 (m, 2H), 2.58 (s, 3H), 1.97-1.91 (m, 3H), 1.91-1.87 (m, 6H), 1.86-1.82 (m, 1H), 1.81-1.76 (m, 9H). $^13$C NMR (101 MHz, CDCl$_3$) $\delta$~197.87, 156.43, 134.63, 128.26, 125.35, 43.69, 38.07, 37.68, 36.62, 34.84, 26.52, 25.65. IR (neat) $\nu_{max}=$2911, 2884, 2848, 1679, 1605, 1268, 1048 cm$^{-1}$. HRSM (DART) calcd. for [C$_{22}$H$_{26}$O$+$H]$^+$ 307.20564, found 307.20560.

\smallskip %
\noindent\textit{4-(4-diamantyl)phenylacetate}: Under open atmosphere, diamantane precursor (176 mg, 0.58 mmol), m-CPBA (296 mg, 1.73 mmol), and NaH$_2$PO$_4$ (276 mg, 2.3 mmol) were suspended in 4 mL DCE. The mixture was stirred at 50$^{\circ}$C for 6 hr, at which point TLC (silica, 95/5 hexane/ethyl acetate) showed all the starting material depleted. The reaction mixture was cooled to room temperature and filtered. The filtrate was diluted with DCM and washed with sat. Na$_2$S$_2$O$_3$, sat. NaHCO$_3$, water, brine and dried over Na$_2$SO$_4$. Solvent was removed under vacuum to afford a yellow solid. The crude material was purified on silica (9/1 hexane/ethyl acetate) to give a white crystalline solid. Yield: 169 mg (90\%). m.p. 137-140$ ^\circ C$. $R_f=0.40$ (silica, 10/1 hexanes/ethyl acetate). $^1$H NMR (400 MHz, CDCl$_3$) $\delta$~7.40-7.33 (m, 2H), 7.05-6.98 (m, 2H), 2.29 (s, 3H), 1.91 (m, 3H), 1.89-1.84 (m, 6H), 1.84-1.80 (m, 1H), 1.79-1.73 (m, 5H). $^13$C NMR (101 MHz, CDCl$_3$) $\delta$~169.70, 148.34, 148.27, 126.16, 120.85, 44.04, 38.22, 37.74, 36.68, 34.10, 25.73, 21.15. IR (neat) $\nu_{max}=$2908, 2872, 2848, 1763, 1508, 1367, 1208, 1171  cm$^{-1}$. HRSM (DART) calcd. for [C$_{22}$H$_{24}$O$_2+$H]$^+$ 321.18491, found 321.18491.

\smallskip
\noindent\textit{\textbf{4-(4-diamantyl)phenol}}: Under an open atmosphere, 4-(4-diamantyl)phenyl acetate (110 mg, 0.34 mmol) was suspended in 1 mL MeOH. NaOH (34 mg, 0.85 mmol) in 50 $\mu$L water was added to the suspension, and the reaction stirred for 6 hr at room temperature. The solvent was removed under vacuum, and the residue was suspended in 10mL water and acidified using 6 M HCl. Solid was filtered, washed with water, taken up in ethyl acetate and washed with brine. Ethyl acetate was removed under vacuum, and the solid was passed through a silica plug using 4/1 hexane/acetone to afford a white solid in quantitative yield. m.p. sublimates $>220 ^\circ C$. $R_f=0.15$ (silica, 10/1 hexanes/ethyl acetate). $^1$H NMR (400 MHz, CDCl$_3$) $\delta$~7.28-7.22 (m, 2H), 6.81-6.76 (m, 2H), 4.66 (s, 1H), 1.93-1.89 (m, 3H), 1.86-1.82 (m, 7H), 1.79-1.76 (m, 9H). $^13$C NMR (101 MHz, CDCl$_3$) $\delta$~153.18, 143.26, 126.25, 114.79, 44.20, 38.31, 37.78, 36.73, 33.66, 25.77. IR (neat) $\nu_{max}=$3272, 2906, 2877, 2847, 1704, 1513, 1439, 1244, 826  cm$^{-1}$. HRSM (DART) calcd. for [C$_{20}$H$_{22}$O$+$H]$^+$ 279.17434, found 279.17434.

\subsection{DLIF spectra of SrOPh-x $\Tilde{B}-\Tilde{X}$ transitions and CaOPh-x $\Tilde{A}/\Tilde{B}-\Tilde{X}$ transitions}
Figure \ref{figs:Sr_BX_DLIF} presents the DLIF spectra for the $\Tilde{B}-\Tilde{X}$ transitions of Sr-containing species, while Fig.\ref{figs:Ca-DLIF} shows all spectra of Ca-containing species. All peaks are assigned by comparing the peak shifts and relative intensities with the theoretical frequencies (Tables \ref{tab:theo-freq-Ca}-\ref{tab:theo-freq-Sr}) and FCFs (Tables \ref{tab:theo-fcf-caophch3-tub}-\ref{tab:theo-fcf-sroph4diA}) of vibrational decays calculated under harmonic approximation. Tables \ref{table:modes-freq}-\ref{table:sr-vbr} summarize the all vibrational frequencies and intensity fractions for resolved vibrational decays. The vibrational displacements of all resolved fundamental modes are illustrated in Figs. \ref{figs:vib-srophch3}-\ref{figs:vib-caophdiA}.

Figures.\ref{figs:Sr_BX_DLIF}a-b display the $\Tilde{B}-\Tilde{X}$ DLIF spectrum of SrOPh-CH$_3$ and SrOPh-C(CH$_3$)$_3$, respectively. The doublet peaks at approximately 200 cm$^{-1}$ in Fig. \ref{figs:Sr_BX_DLIF}a  indicate the Fermi resonance coupling between the $5_1$ and $2_14_1$ modes in the ground state \cite{zhu2024extending}, which is also observed in the $\Tilde{A}-\Tilde{X}$ spectra shown in Fig. \ref{fig1:sroph-AX}(a).  
Figs. \ref{figs:Sr_BX_DLIF}c-e present the dispersed spectra for SrOPh-Ad,  SrOPh-1diA and SrOPh-4diA, respectively, originating from the 
$\widetilde B$ state with excitations at around 658 nm. Similarly, the Fermi resonance effect is also observed in these molecules, as indicated by the doublets near the stretching mode transitions $6_0^1$ at around $150-170$ cm$^{-1}$ in these spectra. The spectrum shown in Fig. \ref{figs:Sr_BX_DLIF}e contains a Ca atomic line (labeled with $+$).
In addition, all spectra in Fig. \ref{figs:Sr_BX_DLIF} show vibrational decays from the $\Tilde{A}$ levels.
These decays are likely due to the collisional relaxation \cite{zhu2022functionalizing,lao2022sroph} or vibronic couplings between the $\Tilde{A}$ and $\Tilde{B}$ states.

For Ca species, the dispersed spectra of CaOPh-CH$_3$ and CaOPh-C(CH$_3$)$_3$ are shown in Figs. \ref{figs:Ca-DLIF}a-d, where most of the intensity ratios for the observed off-diagonal decays match well with the theoretical prediction indicated by blue sticks. 
Due to the coincident excitations, the background signals from CaOH and CaOCH$_3$ with relative intensity around $10^{-2}-10^{-3}$ were observed, which are recognized by the comparison to the data reported in Ref.\cite{bernath1985spectroscopy, crozet2002a2e, brazier1986laser} and the broader line shapes (FWHM = $20-40$cm$^{-1}$) compared to the lines that can be well assigned to the CaOPh-CH$_3$ and CaOPh-C(CH$_3$)$_3$ transitions. The calcium atomic lines at 616 and 657 nm were also observed due to the ablation of the Ca/CaH$_2$ targets. 
Figs. \ref{figs:Ca-DLIF}e-j show the DLIF spectra of the Ca-containing molecules with diamondoids as the ligand. The observed lines can be reasonably assigned to either the atomic lines or the vibrational decays of the molecules. Compared to the Sr-containing molecules, a higher number of off-diagonal vibrational decays were observed in the DLIF spectra of the Ca-containing molecules.  
Additionally, similar to the results of the Sr-containing molecules, neither the number of the observed off-diagonal decays nor the diagonal vibrational branching fractions showed a systematic correlation with the size of the hydrocarbon ligand. 

In general, the frequencies of observed vibrational modes in these molecules match very well with the corresponding results from $\Tilde{A}-\Tilde{X}$ scans, and the differences between the two measurements do not exceed the systematic uncertainty induced by the spectrometer and ICCD camera ($\approx 3$ cm$^{-1}$), see Table. \ref{table:modes-freq}. The measured frequencies of the fundamental modes are calculated from the average values of the observations from the $\Tilde{A}-\Tilde{X}$ and $\Tilde{B}-\Tilde{X}$ scans, and their uncertainties are evaluated from the differences in the observations and the aforementioned uncertainty of the experimental setup. 

The comparison of the measured intensity ratios with the theoretical results can be found in Tables. \ref{table:ca-vbr}-\ref{table:sr-vbr}. Though there are some discrepancies ($<8\%$) in the calculated and measured diagonal VBRs, the strongest few off-diagonal transitions can be well predicted by the calculation in terms of the orders of magnitude, which therefore provides a good reference for the peak assignments. 

\subsection{Theoretical methods}
All the DFT electronic structure calculations were performed using the Gaussian16 package \cite{frisch2016gaussian}. We employed the PBE0-D3/def2-TZVPPD level of theory, consistent with our previous works.
All the excited state optimizations and frequency calculations were done with the excited state gradients produced by the TDDFT method. 
For the estimation of the anharmonicity and treatment of the accidental vibrational mode degeneracies in the explored molecules we perfomed the second order Vibrational Perturbation Theory (VPT2) calculations using the potential expansions obtained at the DFT level, as implemented in the Psience python package.\cite{VPT2-1, zhu2024extending} 
Frank-Condon factors used in the VPT calculations were calculated using ezFCF software.\cite{ezfcf}

Multi-reference calculations of the vertical excitation energies (VEEs) were performed using the Molpro software package.\cite{molpro, 10.1063/1.5094644, Angeli2007} Table \ref{tab:vee-molpro} shows computed VEEs for several systems investigated in this work at the NEVPT2(9,10) level of theory with the same basis set as the DFT/TDDFT calculations.

Density Functional Theory calculations on periodic structures (\textit{i.e.} diamond surfaces) were performed with VASP\cite{RN12}, version 5.4.4, using the PAW\cite{RN14}-PBE\cite{RN13} method with plane-wave cutoff of 600 eV and a spin-unrestricted (UKS) formalism.
The bulk unit cell of diamond was optimized with a 8-8-8 $k$-point grid until the components of the forces on all atoms were less than 0.01 eV/\AA{}.
A (111) surface was constructed with an in-plane lattice parameter of 10 \AA{} and a thickness of 3 atomic layers, and terminated with hydrogen atoms at the upper and lower faces.
The hydrogen positions of this slab were optimized, keeping the carbon atoms fixed, with a 2-1-2 $k$-point grid ($a$ and $c$ are the in-plane lattice vectors).
One hydrogen atom was then replaced by a phenyl group and the structure was reoptimized, with the phenyl group and 1 carbon atom of the surface relaxed and all other atoms fixed.
Finally, the uppermost hydrogen atom of the phenyl group was replaced by OCa and the optimization was repeated.
The optimized C-O bond length was 1.344 \AA{} and the optimized O-Ca bond length was 1.980 \AA{}.
Charge densities for individual orbitals were evaluated using pymatgen and visualised with VESTA.
Phonons were computed for the PhOCa group and the nearest 4 C atoms of the surface at the $\Gamma$ point with the finite differences method, and vibrational modes were visualised using VaspVib2XSF and VESTA.

Highly accurate wavefunction-based electronic structure calculations, particularly of excited states, are generally difficult for periodic systems due to the high scaling of the computational cost of such methods with system size.
This is particularly true for defects and adsorbed molecules, of interest for optical cycling and other quantum information applications, due to large supercells being required to avoid artificial interactions between periodic images. To circumvent this, some of the authors have devised an embedding approach based on the $^\prime$aperiodic defect model$^\prime$,\cite{lavroff2024aperiodic} in which a pristine non-defective crystal is used for the calculation of the embedding field, while the atoms of a fragment, subjected to this field, can be afterward manipulated to create the defect. The fragment is then treated using any appropriate quantum chemical theory including multireference.

\begin{figure}
    \centering
    \includegraphics[width=1\linewidth]{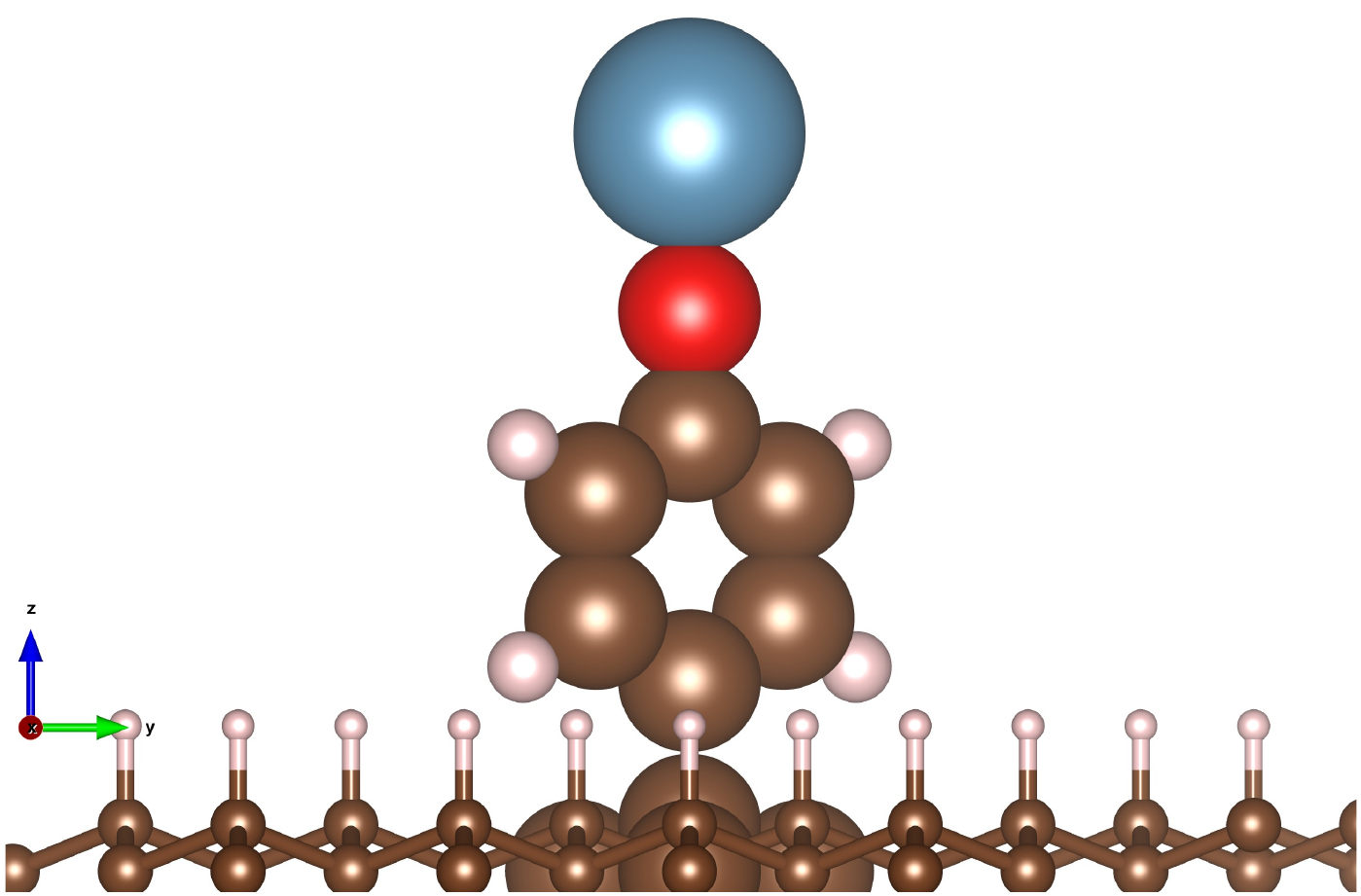}
    \caption{Visualization of the embedded fragment and the nearby atoms of its periodic surroundings. Space-filling atoms are part of the embedded fragment, while ball-and-stick atoms are part of its environment. Carbon atoms are in brown, hydrogen atoms in white, oxygen atoms in red, and calcium atoms in blue.}
    \label{fig:Rob_Embedding}
\end{figure}

\begin{figure*}
\centering
    \includegraphics[scale=0.35]{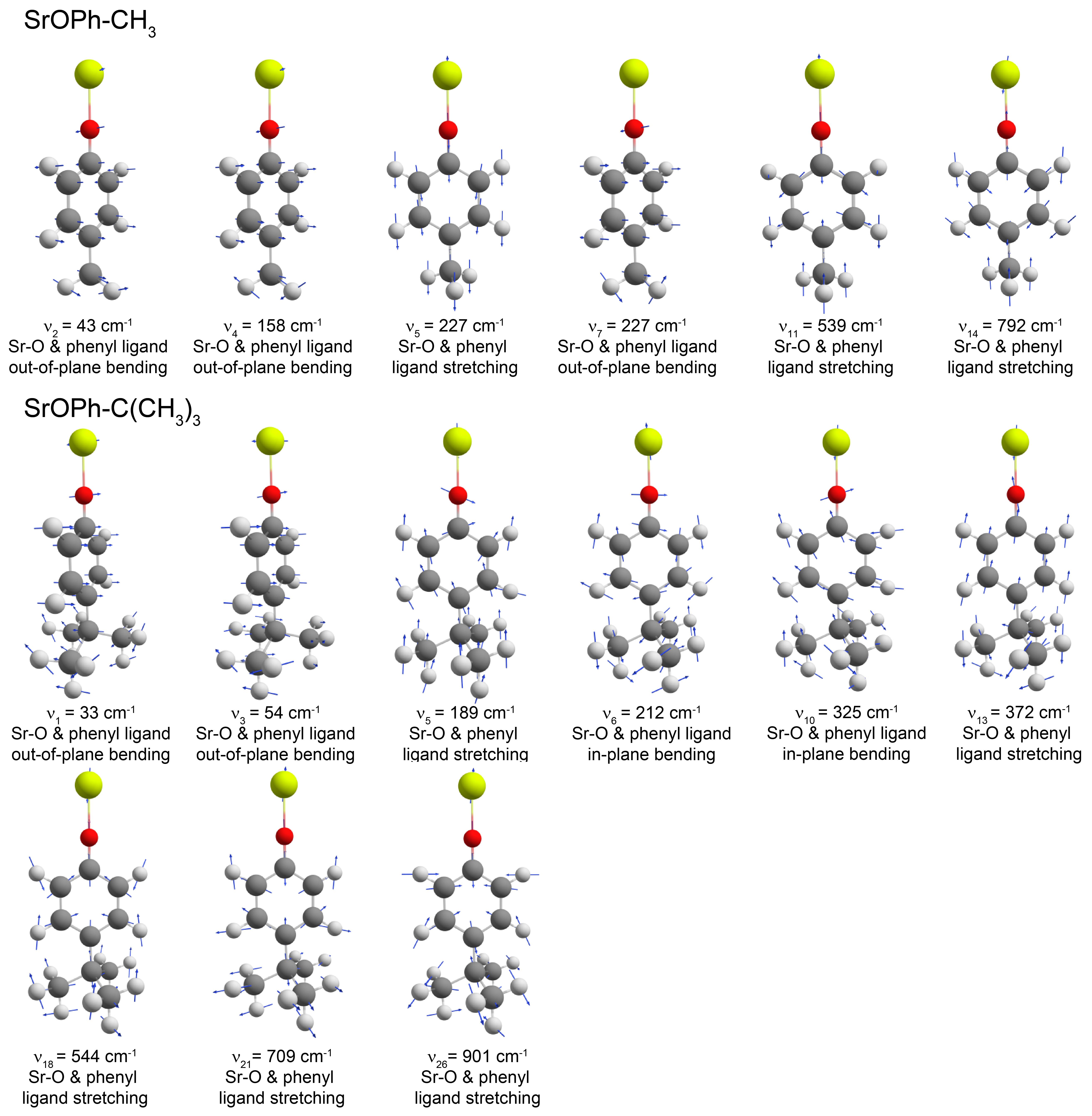}
    \caption{Vibrational displacements of related fundamental vibrational modes for SrOPh-CH$_3$ and SrOPh-C(CH$_3$)$_3$. The calculated harmonic frequencies are given.}
    \label{figs:vib-srophch3}
\end{figure*}

\begin{figure*}
\centering
    \includegraphics[scale=0.35]{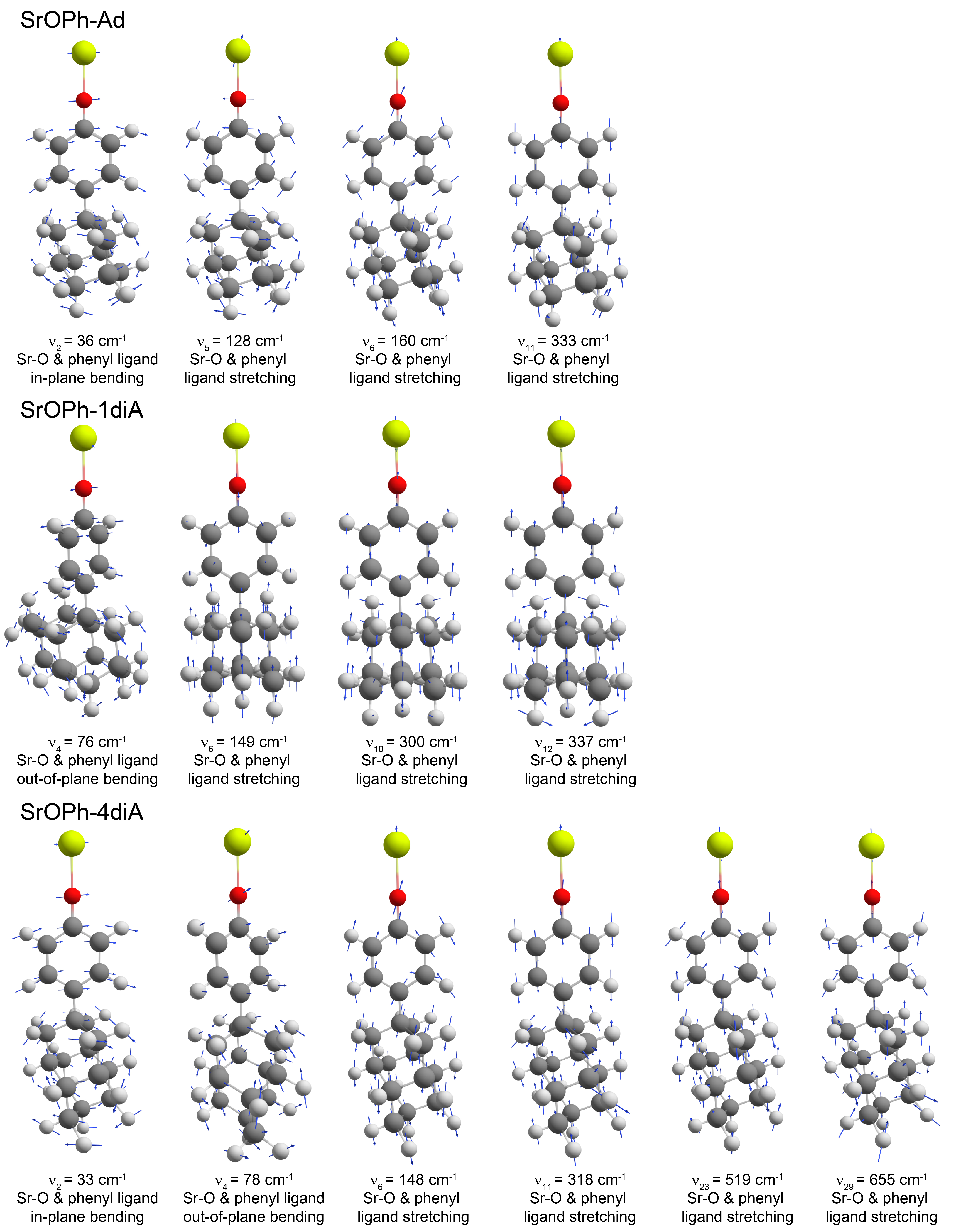}
    \caption{Vibrational displacements of related fundamental vibrational modes for SrOPh-Ad, SrOPh-1diA and SrOPh-4diA. The calculated harmonic frequencies are given.}
    \label{figs:vib-srophAd}
\end{figure*}

\begin{figure*}
\centering
    \includegraphics[scale=0.35]{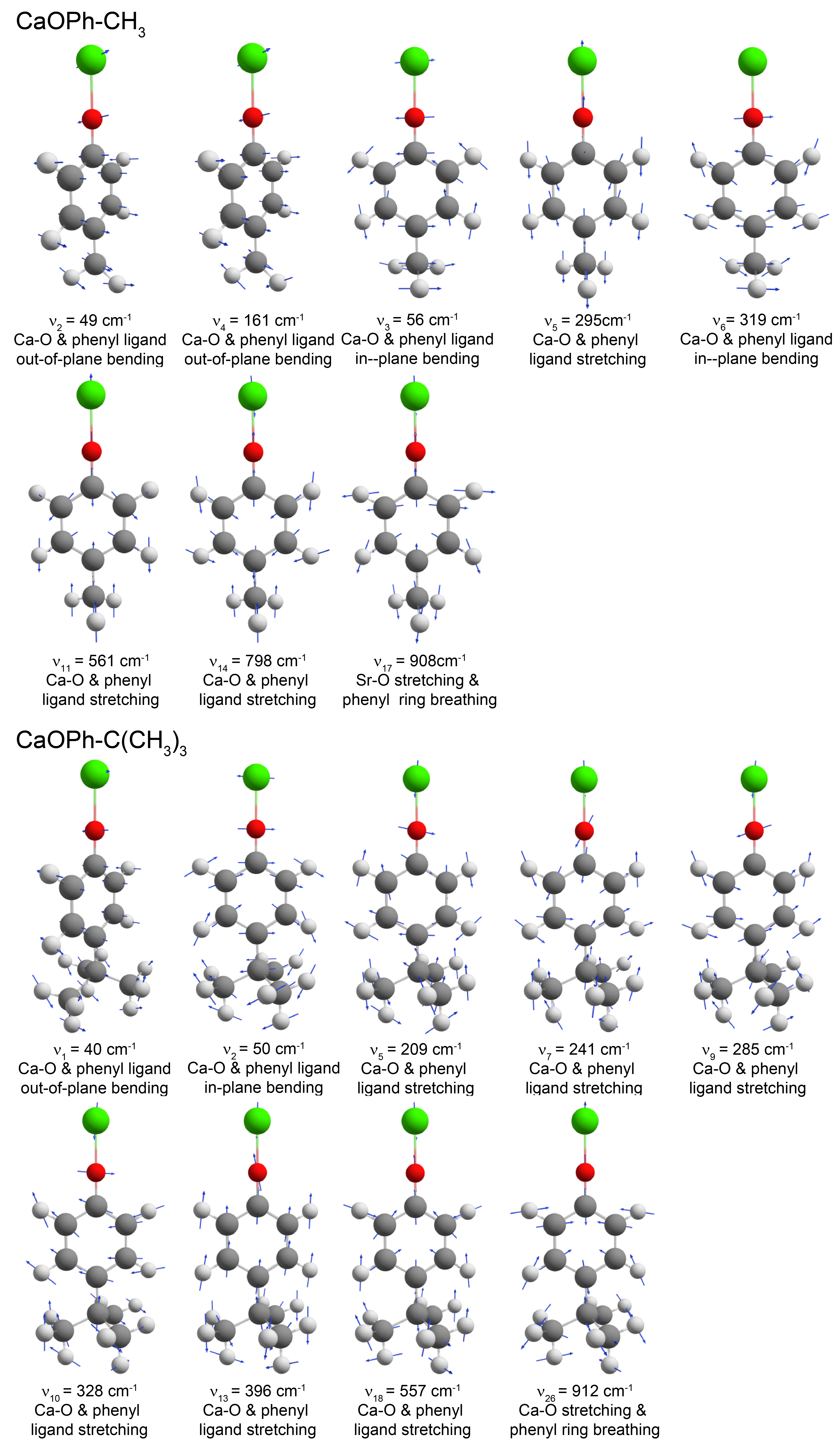}
    \caption{Vibrational displacements of related fundamental vibrational modes for CaOPh-CH$_3$ and CaOPh-C(CH$_3$)$_3$. The calculated harmonic frequencies are given.}
    \label{figs:vib-caophch3}
\end{figure*}

\begin{figure*}
\centering
    \includegraphics[scale=0.35]{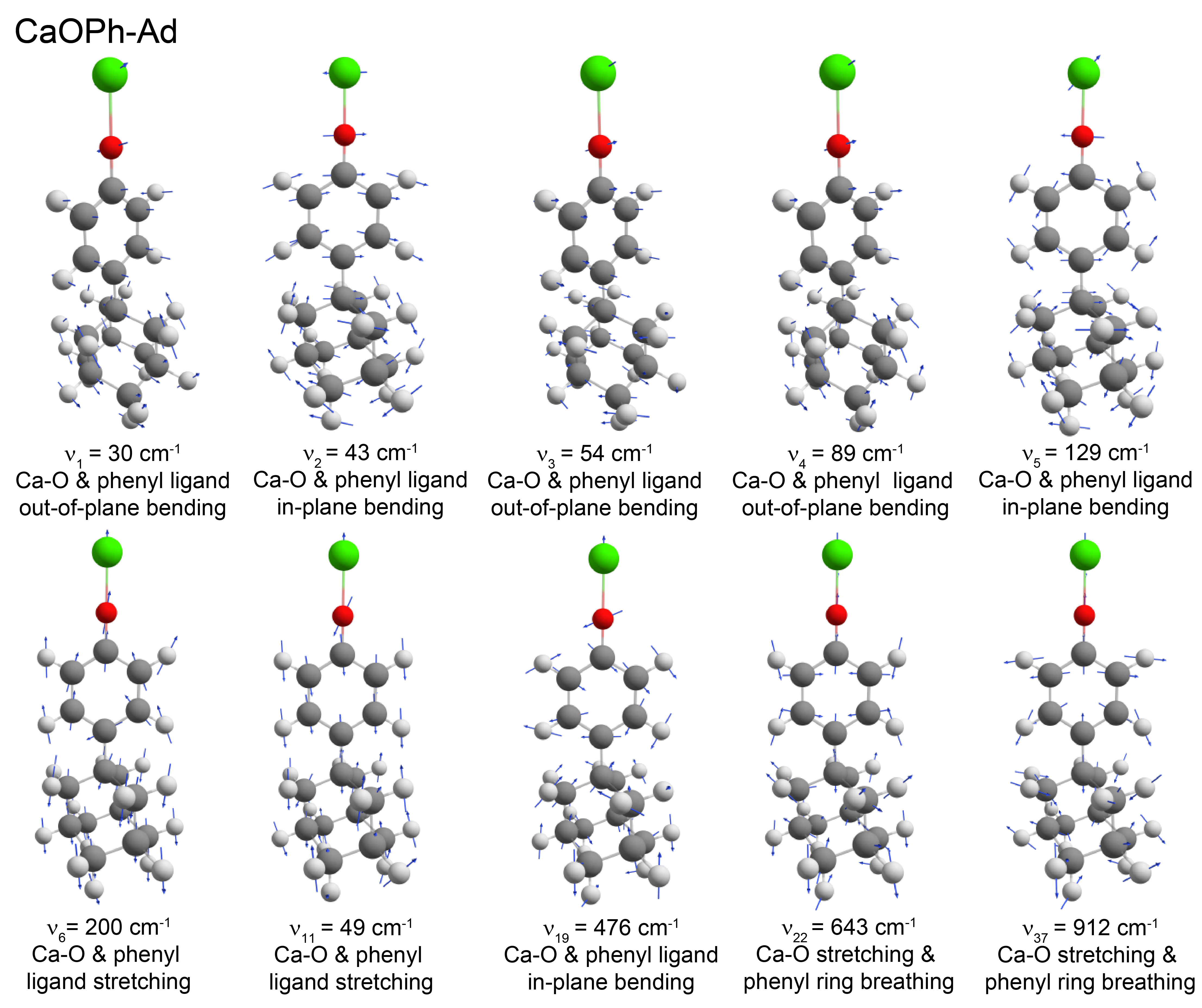}
    \caption{Vibrational displacements of related fundamental vibrational modes for CaOPh-Ad. The calculated harmonic frequencies are given.}
    \label{figs:vib-caophAd}
\end{figure*}

\begin{figure*}
\centering
    \includegraphics[scale=0.35]{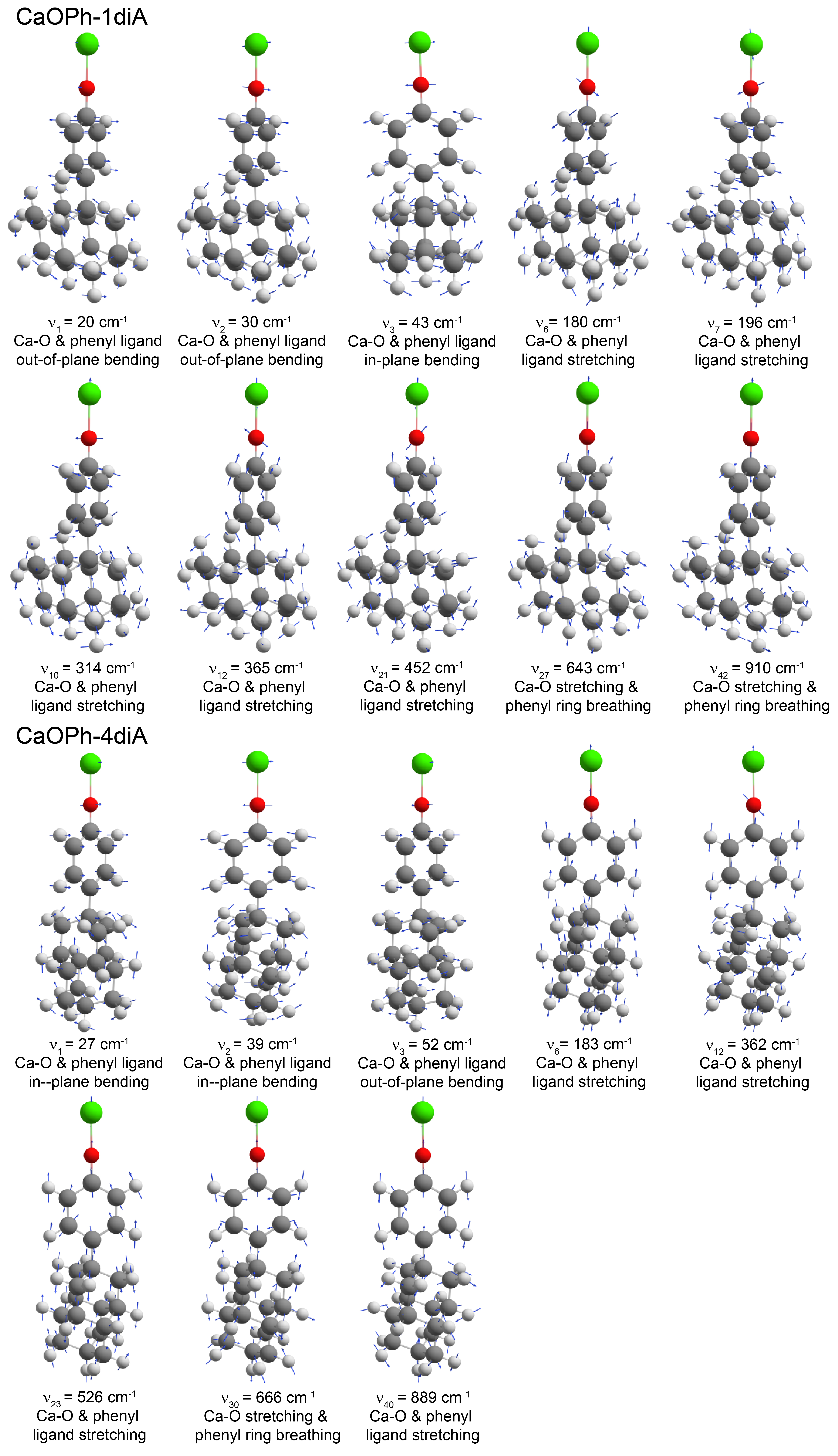}
    \caption{Vibrational displacements of related fundamental vibrational modes for CaOPh-1diA and CaOPh-4diA. The calculated harmonic frequencies are given.}
    \label{figs:vib-caophdiA}
\end{figure*}

\begin{figure*}
\centering
    \includegraphics[scale=0.35]{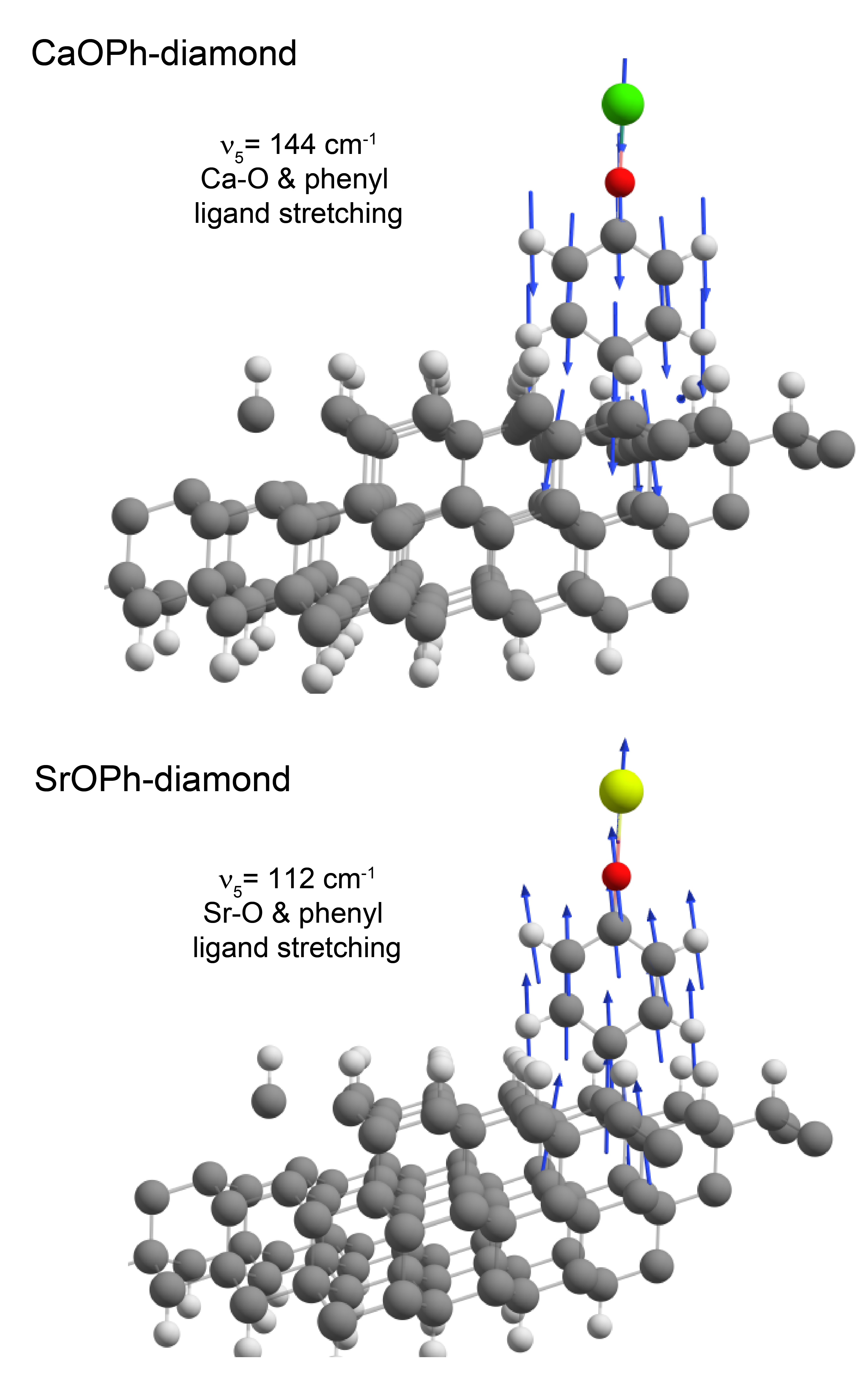}
    \caption{Vibrational displacements of the lowest-frequency stretching vibrational modes for CaOPh-diamond and SrOPh-diamond. The calculated harmonic frequencies are given.}
    \label{figs:vib-diamond}
\end{figure*}

To mimic the diamond surface we employed 
a pristine diamond (111) slab, 5 layers thick and terminated with hydrogen atoms (with a DFT-computed C-H bond length of 1.11 Angstroms). The defect was created by replacing one surface hydrogen with the calcium phenoxide moiety (Figure \ref{fig:Rob_Embedding}).
Because forces and geometry optimizations are not yet available in the aperiodic defect model within the Cryscor code 
\cite{pisani2012,usvyat2010}, we used the aforementioned DFT bond lengths and angles computed in VASP. Within the aperiodic defect approach, \cite{lavroff2024aperiodic} the fragment Hamiltonian, which includes the altered geometry of the defect and the Hartree-Fock embedding field from the pristine crystalline environment, was passed to a molecular program (in this case Molpro \cite{molpro} or pySCF \cite{Sun2020}) using the FCIDUMP file \cite{Knowles89} interface, for the quantum chemical treatment.
The  def2-TZVPPD basis was used for Ca and O, and pob-VTZP-rev2\cite{VilelaOliveira2019} for C and H. Both used the density fitting basis set optimized for MP2/cc-pVTZ.\cite{weigend02b}

We computed vertical excitation energies of the embedded fragment, consisting of CaOPh and its four closest diamond carbon atoms, at the second-order N-electron valence perturbation theory (NEVPT2)\cite{Angeli2007} level. The reference states were calculated using the state averaged complete active space self-consistent field (SA-CASSCF) \cite{10.1063/1.5094644} with an active space of 9 electrons in 10 orbitals.
The NEVPT2 VEEs for the lowest two states are 1.926 eV and 1.947 eV, respectively. These values also agree up to several meV with unrestricted EOM-CCSD, calculated with pySCF \cite{Sun2020} via the same FCIDUMP interface: 1.929 eV and 1.950 eV.

\subsection{Error analysis of VBRs}
All observed peaks in DLIF spectra are fitted with the Voigt function in PGopher, with line intensities estimated from the areas under these fitted curves. The statistical uncertainties of the fitting parameters are estimated with the covariance matrix. The Jacobian conversion from wavelength and frequency is performed as a correction to the calculation of line intensities \cite{mooney2013get}. In addition, as the spectra were taken in a wide wavelength range (610-750nm) while the ICCD sensor quantum efficiency (QE) curve, $>40\%$ in range of 600-615nm, drops to below $20\%$ as wavelength $>$ 710nm (the QE data was provided by the vendor, Oxford Instrument - Andor Technology), the scaling with the inverse of quantum efficiency is therefore applied to the line intensity calculation. The two corrections, denoted as $\Delta_J$ and $\Delta_{QE}$, respectively, are calculated from the differences between the intensity ratios before and after the scaling. The corrections can be larger than the statistical uncertainties, hence, we listed these correction terms for the diagonal VBRs in the table \ref{DiagonalVBRs}, for comparison. 

The unobserved peaks which contribute to the VBRs can lead to system uncertainties \cite{zhu2022functionalizing}, as the true VBRs depend on contributions of all possible decay pathways. Due to the limitations of the measurement sensitivity and the detection window, only a few vibrational decays have been observed for each transition. Compared to a complete description of vibrational decays obtained from calculated FCFs, all unobserved vibrational decays are therefore a source of the systematic uncertainty, which is estimated by \cite{zhu2022functionalizing}: 
\begin{equation}
\label{ScaledVBRs}
\begin{aligned}
  S'_{0}=\frac{S_{0}^{(c)}}{\sum_{\substack{i=0}}^{p}S_{i}^{(c)} + \sum_{\substack{i=p+1}}^{N}\frac{T_{i}}{C}},
\end{aligned}
\end{equation}
where $p$ is the number of observed vibrational decays, $T_i$ is the theoretical VBRs, $C=\frac{1}{p+1}\sum_{i=0}^p\frac{T_i}{S_{i}^{(c)}}$ is a scaling factor, and $S_{i}^{(c)}=S_{i}+\Delta_{J, i}+\Delta_{QE, i}$ denotes the intensity ratios with corrections. For a straight forward comparison between the scaled VBRs, $S'_{i}$, and the intensity ratios, $S_{i}^{(c)}$, we denote their differences as $\delta_{u, i} = S'_{i}-S_{i}^{(c)}$. The values of $\delta_u$ for the diagonal peaks can be found in Table.\ref{DiagonalVBRs}. As $S_{0}^{(c)}$ is always greater than $S_{0}'$, it is regarded as the estimates for the upper bound of the actual diagonal VBRs, and $S_{0}'$ is the reported value for the diagonal VBRs in this work. To estimate the lower bound of the VBRs, $S_{0}''$, we choose a smaller scaling factor $C'=C-\delta_C$ to replace $C$ in Eq. \ref{ScaledVBRs}, where $\delta_C$ is the uncertainty of the scaling factor: 
\begin{equation}
\begin{split}
    \delta_C=&\sqrt{\sum_{i=0}^p\left(\frac{\partial C}{\partial S_{i}^{(c)}}\delta S_{i}^{(c)}\right)^2+\left(\frac{\partial C}{\partial p}\times 1\right)^2}\\
    \approx&\sqrt{\sum_{i=0}^pT_i^2\left(\frac{\delta S_{i}^{(c)}}{(p+1)S_{i}^{(c)}}\right)^2+\left(\frac{1-C}{p+2}\right)^2},
\end{split}
\end{equation} 
here $\delta S_{i}^{(c)}$ is the uncertainty of $S_{i}^{(c)}$. In the last step, we
assumed that $\frac{T_{p+1}}{S_{p+1}^{(c)}}\approx 1$ and calculated the $\frac{\partial C}{\partial p}$ from a direct assumption that $\frac{\partial C}{\partial p}\approx \frac{1}{p+2}\sum_{i=0}^{p+1}\frac{T_i}{S_{i}^{(c)}}-\frac{1}{p+1}\sum_{i=0}^{p}\frac{T_i}{S_{i}^{(c)}}\approx \frac{p+1}{p+2}C+\frac{1}{p+2}-C$.

\subsection{Functionalization of diamond surfaces}
\label{sec:surface occ}
MOPh-diamond can be synthesized through chemical functionalization on hydrogen-terminated diamond surfaces.~\cite{2008diamondfunction,2019diamondfunction}. As shown in Scheme \ref{scheme2}, three procedures will be used to produce (CaOPh)$_n$-diamond complex. A crucial step, labeled as Step 2, involves chemically bonding phenol ligands onto the diamond surface, typically terminated with hydrogen atoms. This bonding process can be achieved by electrochemical grafting of freshly grown diamonds or boron-doped diamonds\cite{diamond2015doping} with phenol-based diaonium salt. The diazonium salt can be synthesized via diazotization of 4-aminophenol in Step 1. Upon formation of the phenol layers on diamond surfaces, metastable calcium atoms, generated either from a hot oven or laser ablation, are introduced to react with the phenol-diamond complex in a vacuum environment, leading to the formation of (CaOPh)$_n$-diamond. Using various chemical treatments, the terminated hydrogen atoms on the diamond surface can be substituted with oxygen, nitrogen and sulfur atoms. This allows for the functionalization of the diamond surface with a wide range of bond types and ligands to suit specific applications.

\begin{scheme*}
    \centering
    \includegraphics[scale=0.8]{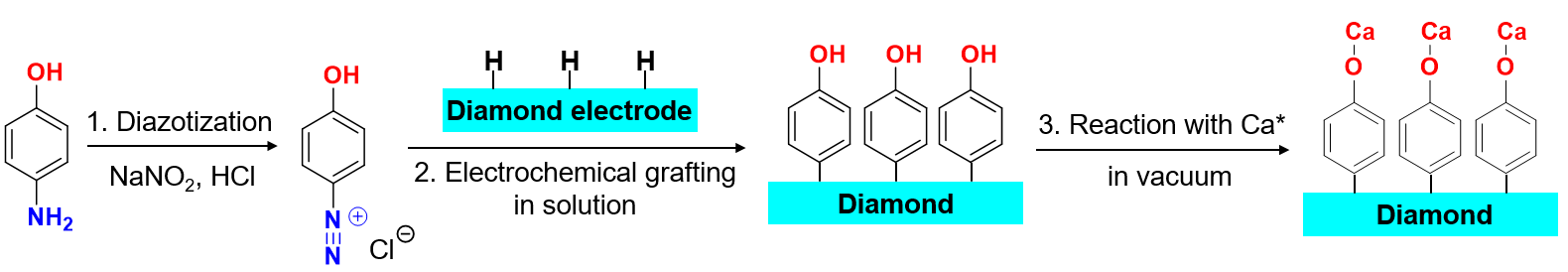}
    \caption{Procedures to synthesize (CaOPh)$_n$-diamond from 4-aminophenol and H-terminated diamond surface. A boron doped diamond with lower electrical resistivity can be used to improve the efficiency of the electrochemical diazonium grafting process.
    }
   \label{scheme2}
\end{scheme*}

\renewcommand*{\thetable}{S\arabic{table}}
\setcounter{table}{0}

\begin{table*}
    \centering
\begin{tabular}{ccc}
    \hline
    \hline
    Molecules &  X $\rightarrow$ A  & X $\rightarrow$ B   \\
    \hline
    CaOPh-CH$_3$             & 1.7639 &	1.7878  \\
    CaOPh-C(CH$_3$)$_3$      & 1.7639 &	1.7870  \\
    SrOPh-CH$_3$             & 1.6832 &	1.7004  \\
    CaOPh-diamantane         & 1.8983 &	1.8983  \\
    SrOPh-diamantane         & 1.9014 &	1.9171  \\
    \hline
    \hline
    \end{tabular}
    \caption{Vertical excitation energies for several molecules calculated using NEVPT2(9,10) level of theory reported in eV. }
    \label{tab:vee-molpro}
\end{table*}

\begin{table*}[] 
\centering
\caption{Calculated harmonic frequencies of all Ca-species. A full list of vibrational modes of CaOPh-CH$_3$ and the lowest fifty modes of all other molecules are presented. All vibrational modes are labeled with increasing frequency.}
\label{tab:theo-freq-Ca}
\setlength{\tabcolsep}{7pt}
\renewcommand{\arraystretch}{1.2}
\begin{tabular}{cccccccccc}
\hline
\hline
\multicolumn{2}{l}{CaOPh-CH3} & \multicolumn{2}{l}{CaOPh-C(CH3)3} & \multicolumn{2}{l}{CaOPh-Ad} & \multicolumn{2}{l}{CaOPh-1diA} & \multicolumn{2}{l}{CaOPh-4diA} \\ \hline
Modes & Freq. & Modes & Freq. & Modes & Freq. & Modes & Freq. & Modes & Freq. \\ \hline
$\nu_1$ & 37 & $\nu_1$ & 40 & $\nu_1$ & 30 & $\nu_1$ & 20 & $\nu_1$ & 27 \\
$\nu_2$ & 49 & $\nu_2$ & 50 & $\nu_2$ & 43 & $\nu_2$ & 31 & $\nu_2$ & 39 \\
$\nu_3$ & 56 & $\nu_3$ & 54 & $\nu_3$ & 54 & $\nu_3$ & 45 & $\nu_3$ & 52 \\
$\nu_4$ & 161 & $\nu_4$ & 113 & $\nu_4$ & 89 & $\nu_4$ & 78 & $\nu_4$ & 81 \\
$\nu_5$ & 295 & $\nu_5$ & 209 & $\nu_5$ & 129 & $\nu_5$ & 130 & $\nu_5$ & 106 \\
$\nu_6$ & 319 & $\nu_6$ & 228 & $\nu_6$ & 200 & $\nu_6$ & 180 & $\nu_6$ & 183 \\
\hline
\hline
\end{tabular}
\end{table*}

\begin{table*}[]
\centering
\caption{Calculated harmonic frequencies of all Sr-species. A full list of vibrational modes of SrOPh-CH$_3$ and the lowest fifty modes of all other molecules are presented. All vibrational modes are labeled with increasing frequency.}
\label{tab:theo-freq-Sr}
\setlength{\tabcolsep}{7pt}
\renewcommand{\arraystretch}{1.2}

\begin{tabular}{cccccccccc}
\hline
\hline
\multicolumn{2}{c}{SrOPh-CH3} & \multicolumn{2}{c}{SrOPh-C(CH3)3} & \multicolumn{2}{c}{SrOPh-Ad} & \multicolumn{2}{c}{SrOPh-1diA} & \multicolumn{2}{c}{SrOPh-4diA} \\ \hline
Modes & Freq. & Modes & Freq. & Modes & Freq. & Modes & Freq. & Modes & Freq. \\ \hline
$\nu_1$ & 36 & $\nu_1$ & 33 & $\nu_1$ & 26 & $\nu_1$ & 20 & $\nu_1$ & 23 \\
$\nu_2$ & 43 & $\nu_2$ & 44 & $\nu_2$ & 36 & $\nu_2$ & 26 & $\nu_2$ & 33 \\
$\nu_3$ & 51 & $\nu_3$ & 54 & $\nu_3$ & 54 & $\nu_3$ & 37 & $\nu_3$ & 51 \\
$\nu_4$ & 158 & $\nu_4$ & 109 & $\nu_4$ & 86 & $\nu_4$ & 76 & $\nu_4$ & 77 \\
$\nu_5$ & 227 & $\nu_5$ & 189 & $\nu_5$ & 128 & $\nu_5$ & 130 & $\nu_5$ & 105 \\
$\nu_6$ & 318 & $\nu_6$ & 212 & $\nu_6$ & 160 & $\nu_6$ & 149 & $\nu_6$ & 148 \\
\hline
 \hline
\end{tabular}
\end{table*}

\begin{table*}
\centering
\setlength{\tabcolsep}{6pt}
\caption{Theoretical harmonic frequencies and FCFs ($\gtrsim 10^{-4}$) of vibrational decays from A-X and B-X transitions for CaOPh-CH3 and CaOPh-tbu. VBRs are derived from calculated FCFs of corresponding vibrational decays.}
\label{tab:theo-fcf-caophch3-tub}
\begin{tabular}{ccccccccccccc}
\hline
\hline
\multicolumn{6}{c}{CaOPh-CH3}                                       &                      & \multicolumn{6}{c}{CaOPh-tbu}                                       \\ \cline{1-6} \cline{8-13} 
\multicolumn{3}{c}{\textbf{A-X}} & \multicolumn{3}{c}{\textbf{B-X}} &                      & \multicolumn{3}{c}{\textbf{A-X}} & \multicolumn{3}{c}{\textbf{B-X}} \\ \cline{1-6} \cline{8-13} 
Final     & Harm.       &          & Final       & Harm.   &          & \multicolumn{1}{l}{} & Final       & Harm.   &          & Final       & Harm.   &          \\
states    & freq.       & FCFs      & states      & freq.   & FCFs      &                      & states      & freq.  & FCFs      & states      & freq.   & FCFs      \\ \cline{1-6} \cline{8-13} 
$0$       & 0           & 0.9524 & $0$       & 0           & 0.9695 &  & $0$       & 0           & 0.9099 & $0$       & 0           & 0.9683 \\
$5_1$     & 295         & 0.0342 & $5_1$     & 295         & 0.0213 &  & $2_2$     & 100         & 0.0321 & $7_1$     & 241         & 0.0139 \\
$11_1$    & 561         & 0.0052 & $11_1$    & 561         & 0.0047 &  & $7_1$     & 241         & 0.0213 & $13_1$    & 396         & 0.0075 \\
$17_1$    & 908         & 0.0023 & $17_1$    & 908         & 0.0014 &  & $13_1$    & 396         & 0.0111 & $18_1$    & 557         & 0.0027 \\
\hline
\hline
\end{tabular}
\end{table*}

\begin{table*}
\centering
\setlength{\tabcolsep}{6pt}
\caption{Theoretical harmonic frequencies and FCFs ( $\gtrsim 10^{-4}$) of vibrational decays from A-X and B-X transitions for CaOPh-Ad and CaOPh-1diA. VBRs are derived from calculated FCFs of corresponding vibrational decays.}
\label{tab:theo-fcf-caophAd-1diA}
\begin{tabular}{ccccccccccccc}
\hline
\hline
\multicolumn{6}{c}{CaOPh-ad}                                        &  & \multicolumn{6}{c}{caoph-1diA}                                      \\ \cline{1-6} \cline{8-13} 
\multicolumn{3}{c}{\textbf{A-X}} & \multicolumn{3}{c}{\textbf{B-X}} &  & \multicolumn{3}{c}{\textbf{A-X}} & \multicolumn{3}{c}{\textbf{B-X}} \\ \cline{1-6} \cline{8-13} 
Final        & Harm.  &          & Final        & Harm.  &          &  & Final        & Harm.  &          & Final       & Harm.   &          \\
states       & freq.  & FCFs      & states       & freq.  & FCFs      &  & states       & freq.  & FCFs      & states      & freq.   & FCFs      \\ \cline{1-6} \cline{8-13} 
$0$          & 0      & 0.9272  & $0$          & 0      & 0.9457  &  & $0$          & 0      & 0.9260  & $0$         & 0       & 0.9516  \\
$6_1$        &  200   & 0.0208  & $6_1$        &  200   & 0.0130  &  & $12_1$       &  365   & 0.0181  & $12_1$      &  365    & 0.0120  \\
$11_1$       &  374   & 0.0167  & $3_2$        &  109   & 0.0112  &  & $6_1$        &  180   & 0.0129  & $6_1$       &  180    & 0.0085  \\
$3_12_1$     &  97    & 0.0068  & $11_1$       &  374   & 0.0107  &  & $7_1$        &  195   & 0.0096  & $7_1$       &  195    & 0.0060  \\
\hline
\hline
\end{tabular}
\end{table*}

\begin{table*}
\centering
\setlength{\tabcolsep}{6pt}
\caption{Theoretical harmonic frequencies and FCFs ($\gtrsim 10^{-4}$) of vibrational decays from A-X and B-X transitions for CaOPh-4diA. VBRs are derived from calculated FCFs of corresponding vibrational decays.}
\label{tab:theo-fcf-caoph4diA}
\begin{tabular}{cccccc}
\hline
\hline
\multicolumn{6}{c}{CaOPh-4diA}                                      \\ \hline
\multicolumn{3}{c}{\textbf{A-X}} & \multicolumn{3}{c}{\textbf{B-X}} \\ \hline
Final       & Harm.   &          & Final       & Harm.   &          \\
states      & freq.   & FCFs      & states      & freq.   & FCFs      \\ \hline
$0$         & 0       & 0.9438  & $0$         & 0      & 0.9584  \\
$6_1$       &  183    & 0.0195  & $12_1$      &  62    & 0.0122  \\
$12_1$      &  362    & 0.0191  & $6_1$       &  83    & 0.0120  \\
$23_1$      &  526    & 0.0024  & $3_1$       &  2     & 0.0023  \\
\hline
\hline
\end{tabular}
\end{table*}

\begin{table*}
\centering
\setlength{\tabcolsep}{6pt}
\caption{Theoretical harmonic frequencies and FCFs ($\gtrsim 10^{-4}$) of vibrational decays from A-X and B-X transitions for SrOPh-CH$_3$ and SrOPh-tbu. VBRs are derived from calculated FCFs of corresponding vibrational decays.}
\label{tab:theo-fcf-srophch3-tbu}
\begin{tabular}{ccccccccccccc}
\hline
\hline
\multicolumn{6}{c}{SrOPh-CH3}                                       &  & \multicolumn{6}{c}{SrOPh-tbu}                                       \\ \cline{1-6} \cline{8-13} 
\multicolumn{3}{c}{\textbf{A-X}} & \multicolumn{3}{c}{\textbf{B-X}} &  & \multicolumn{3}{c}{\textbf{A-X}} & \multicolumn{3}{c}{\textbf{B-X}} \\ \cline{1-6} \cline{8-13} 
Final       & Harm.   &          & Final       & Harm.   &          &  & Final       & Harm.   &          & Final        & Harm.  &          \\
states      & freq.   & FCFs      & states      & freq. & FCFs      &  & states      & freq.   & FCFs      & states       & freq.  & FCFs      \\ \cline{1-6} \cline{8-13} 
$0$         & 0       & 0.9221  & $0$         & 0       & 0.9399  &  & $0$         & 0       & 0.9138  & $0$          & 0      & 0.9383  \\
$5_1$       &  227    & 0.0595  & $5_1$       &  227    & 0.0441  &  & $5_1$       &  189    & 0.0496  & $5_1$        &  189   & 0.0368  \\
$1_2$       &  72     & 0.0044  & $1_2$       &  72     & 0.0045  &  & $13_1$      &  372    & 0.0075  & $13_1$       &  372   & 0.0064  \\
$11_1$      &  539    & 0.0041  & $11_1$      &  539    & 0.0041  &  & $6_1$       &  212    & 0.0055  & $6_1$        &  212   & 0.0037  \\
\hline
\hline
\end{tabular}
\end{table*}

\begin{table*}
\centering
\setlength{\tabcolsep}{6pt}
\caption{Theoretical harmonic frequencies and FCFs ($\gtrsim 10^{-4}$) of vibrational decays from A-X and B-X transitions for SrOPh-Ad and SrOPh-1diA. VBRs are derived from calculated FCFs of corresponding vibrational decays.}
\label{tab:theo-fcf-srophAd-1diA}
\begin{tabular}{ccccccccccccc}
\hline
\hline
\multicolumn{6}{c}{SrOPh-Ad}                                        &  & \multicolumn{6}{c}{SrOPh-1diA}                                      \\ \cline{1-6} \cline{8-13} 
\multicolumn{3}{c}{\textbf{A-X}} & \multicolumn{3}{c}{\textbf{B-X}} &  & \multicolumn{3}{c}{\textbf{A-X}} & \multicolumn{3}{c}{\textbf{B-X}} \\ \cline{1-6} \cline{8-13} 
Final        & Harm.  &          & Final       & Harm.   &          &  & Final        & Harm.  &          & Final        & Harm.  &          \\
states       & freq.  & FCFs      & states      & freq.   & FCFs      &  & states       & freq.  & FCFs      & states       & freq.  & FCFs      \\ \cline{1-6} \cline{8-13} 
$0$          & 0      & 0.9097  & $0$         & 0       & 0.9186  &  & $0$          & 0      & 0.8610  & $0$          & 0      & 0.8475  \\
$6_1$        &  160   & 0.0515  & $6_1$       &  160    & 0.0377  &  & $6_1$        &  149   & 0.0450  & $1_2$        &  40    & 0.0541  \\
$11_1$       &  333   & 0.0170  & $11_1$      &  333    & 0.0132  &  & $3_2$        &  75    & 0.0281  & $6_1$        &  149   & 0.0319  \\
$3_12_1$     &  90    & 0.0028  & $22_1$      &  630    & 0.0023  &  & $1_2$        &  40    & 0.0206  & $10_1$       &  300   & 0.0078  \\
\hline
\hline
\end{tabular}
\end{table*}

\begin{table*}
\centering
\setlength{\tabcolsep}{7pt}
\caption{Theoretical harmonic frequencies and FCFs ($\gtrsim 10^{-4}$) of vibrational decays from A-X and B-X transitions for SrOPh-4diA. VBRs are derived from calculated FCFs of corresponding vibrational decays.}
\label{tab:theo-fcf-sroph4diA}
\begin{tabular}{cccccc}
\hline
\hline
\multicolumn{6}{c}{SrOPh-4diA}                                      \\ \hline
\multicolumn{3}{c}{\textbf{A-X}} & \multicolumn{3}{c}{\textbf{B-X}} \\ \hline
Final       & Harm.   &          & Final        & Harm.  &          \\
states      & freq.   & FCFs      & states       & freq.& FCFs      \\ \hline
$0$         & 0       & 0.9144  & $0$          & 0      & 0.9327  \\
$6_1$       &  148    & 0.0502  & $6_1$        &  148   & 0.0376  \\
$11_1$      &  318    & 0.0209  & $11_1$       &  318   & 0.0160  \\
$85_1$      &  1349   & 0.0018  & $3_11_1$     &  74    & 0.0019  \\
\hline
\hline
\end{tabular}
\end{table*}

\begin{table*}
\centering
\caption{The observed frequencies of all resolved vibrational modes in the electronic ground states of all the molecular species involved in this work (unit: cm $^{-1}$). The calculated harmonic frequencies of the fundamental modes are given for comparison. Related vibrational displacements of fundamental modes are presented in Figs. \ref{figs:vib-srophch3}-\ref{figs:vib-caophdiA}. The uncertainties of observed frequencies are within 5 \cm.}
\begin{tabular}{cccccccc|cccccccc}
\hline
\hline
\multicolumn{3}{c}{CaOPh-CH3}  &  & \multicolumn{3}{c}{CaOPh-C(CH3)3} &  &  & \multicolumn{3}{c}{SrOPh-CH3}   &  & \multicolumn{3}{c}{SrOPh-C(CH3)3} \\ \cline{1-3} \cline{5-7} \cline{10-12} \cline{14-16} 
Vib. modes    & Exp.  & Theo.  &  & Vib. modes    & Exp.    & Theo.   &  &  & Vib. modes    & Exp.   & Theo.  &  & Vib. modes    & Exp.    & Theo.   \\ \cline{1-3} \cline{5-7} \cline{10-12} \cline{14-16} 
$2_1$         & 47    & 49     &  & $1_1$         & 38      & 39      &  &  & $2_14_1$      & 197    &        &  & $1_1$         & 30      & 30      \\
$3_2$         & 113   &        &  & $2_1$         & 38      & 39      &  &  & $5_1$         & 226    & 227    &  & $3_2$         & 109     &         \\
$4_1$         & 165   & 161    &  & $1_2$         & 78      &         &  &  & $4_2$         & 313    &        &  & $5_1$         & 187     & 189     \\
$5_1$         & 290   & 295    &  & $5_1$         & 212     & 209     &  &  & $5_2$         & 451    &        &  & $6_1$         & 217     & 212     \\
$6_1$         & 320   & 319    &  & $7_1$         & 240     & 241     &  &  & $11_1$        & 530    & 539    &  & $10_1$        & 321     & 325     \\
$3_1 6_1$     & 365   &        &  & $9_1$         & 275     & 285     &  &  & $14_1$        & 777    & 792    &  & $5_2$         & 375     &         \\
\hline
\hline
\end{tabular}
\label{table:modes-freq}
\end{table*}

\begin{table*}
\centering
\caption{The intensity ratios of all observed vibrational decays for Ca-containing molecules. These are measured from the relative peak areas from the Voigt fits with corrections of ICCD quantum efficiency. The errors indicate the combined statistical uncertainties from the Voigt fits and the ICCD quantum efficiency. The theoretical VBRs are also added for comparison.}
\begin{tabular}{ccccccccccc}
\hline
\hline
\multicolumn{5}{c}{\textbf{CaOPh-CH$_3$}}                              & \textbf{}     & \multicolumn{5}{c}{\textbf{CaOPh-Ad}}                                    \\ \cline{1-5} \cline{7-11} 
Final states & Theo. (A)   & Exp. (A)   & Theo. (B)   & Exp. (B)       &               & Final states & Theo. (A)   & Exp. (A)   & Theo. (B)   & Exp. (B)   \\ \cline{1-5} \cline{7-11} 
$0$          & 0.9562      & 0.9421(13) & 0.9718      & \multicolumn{2}{l}{0.9393(19)} & $0$          & 0.9345      & 0.9247(26) & 0.9503      & 0.9268(19) \\
$2_1$        &             &            & $< 10^{-4}$ & 0.0013(4)      &               & $2_1$        & $< 10^{-4}$ & 0.0043(3)  &             &            \\
$3_2$        &             &            & $< 10^{-4}$ & 0.0032(3)      &               & $3_1$        &             &            & 0.0007      & 0.0042(5)  \\
$4_1$        &             &            & $< 10^{-4}$ & \multicolumn{2}{l}{0.0076(15)} & $1_1 3_1$    & 0.0005      & 0.0030(3)  & 0.0040      & 0.0029(3)  \\
$5_1$        & 0.0325      & 0.0425(12) & 0.0202      & \multicolumn{2}{l}{0.0344(11)} & $2_1 4_1$    & $< 10^{-4}$ & 0.0027(2)  &             &            \\
$6_1$        & $< 10^{-4}$ & 0.0027(3)  & $< 10^{-4}$ & 0.0016(3)      &               & $6_1$        & 0.0202      & 0.0307(19) & 0.0126      & 0.0277(15) \\
\hline
\hline
\end{tabular}
\label{table:ca-vbr}
\end{table*}

\begin{table*}
\centering
\setlength{\tabcolsep}{4pt}
\caption{The intensity ratios of all observed vibrational decays for Sr-containing molecules. These are measured from the relative peak areas from the Voigt fits with corrections of ICCD quantum efficiency. The errors indicate the combined statistical uncertainties from the Voigt fits and the ICCD quantum efficiency. The theoretical VBRs are also added for comparison.}
\begin{tabular}{ccccccccccc}
\hline
\hline
\multicolumn{5}{c}{\textbf{SrOPh-CH$_3$}}                             & \textbf{}     & \multicolumn{5}{c}{\textbf{SrOPh-Ad}}                                    \\ \cline{1-5} \cline{7-11} 
Final states & Theo. (A)   & Exp. (A)  & Theo. (B)   & Exp. (B)       &               & Final states & Theo. (A)   & Exp. (A)      & Theo. (B)      & Exp. (B)   \\ \cline{1-5} \cline{7-11} 
$0$          & 0.9269      & 0.9578(6) & 0.9438      & \multicolumn{2}{l}{0.9161(13)} & $0$          & 0.9156      & 0.9310(12)    & 0.9387         & 0.9162(11) \\
$2_14_1$     & $< 10^{-4}$ & 0.0008(1) & 0.0000      & 0.0039(3)      &               & $5_1$        &             &               & 0.0006         & 0.0023(7)  \\
$5_1$        & 0.0571      & 0.0347(5) & 0.0423      & \multicolumn{2}{l}{0.0701(10)} & $6_1$        & 0.0502      & 0.0380(9)     & 0.0374         & 0.0360(4)  \\
$4_2$        & $< 10^{-4}$ & 0.0008(1) &             &                &               & $2_1 5_1$    & 0.0000      & 0.0142(6)     & $< 10^{-4}$    & 0.0123(4)  \\
$5_2$        & 0.0016      & 0.0014(1) & 0.0008      & 0.0009(2)      &               & $11_1$       & 0.0160      & 0.0168(5)     & 0.0126         & 0.0332(7)  \\ \cline{7-11} 
$11_1$       & 0.0037      & 0.0041(2) & 0.0037      & 0.0077(2)      &               &              &             &               &                &            \\
\hline
\hline
\end{tabular}
\label{table:sr-vbr}
\end{table*}

\begin{table*}
\centering
\caption{Measured intensity ratios and scaled VBRs of the diagonal 0-0 decay of all molecules. The scaling process considering contributions of unobserved vibrational decays is detailed in the section of error analysis of VBRs. The values of CaOPh and SrOPh are obtained from references \cite{zhu2022functionalizing,lao2022sroph,zhu2024extending}.}
\begin{tabular}{cccccccc}
\hline
\hline
\multicolumn{1}{c}{Species} & Transitions & Theo.VBR & Exp. VBR & Exp.VBR errs & Scaled VBR & Scaled VBR & Scaled VBR  \\ 
                            &             &          &          &              &            & upper errs & lower errs \\ \hline
\multirow{2}{*}{CaOPh}      & X-A         & 0.9611   & 0.9339   & 0.0029       & 0.9286     & 0.0053                & 0.0012                \\
                            & X-B         & 0.9758   & 0.9446   & 0.0068       & 0.9339     & 0.0048                & 0.0016                \\
\multirow{2}{*}{CaOPh-CH3}  & X-A         & 0.9562   & 0.9421   & 0.0013       & 0.9357     & 0.0040                & 0.0013                \\
                            & X-B         & 0.9718   & 0.9393   & 0.0019       & 0.9381     & 0.0020                & 0.0019                \\
\multirow{2}{*}{CaOPh-tBU}  & X-A         & 0.9159   & 0.9447   & 0.0012       & 0.9288     & 0.0139                & 0.0020                \\
                            & X-B         & 0.9711   & 0.9389   & 0.0016       & 0.9350     & 0.0027                & 0.0016                \\
\hline
\hline
\end{tabular}
\label{tab:diagonal-VBRs}
\end{table*}

\begin{table*}
    \centering
\begin{tabular}{c|cc|cc|cc|cc|cc|cc}
\hline\hline
 Molecule& \multicolumn{2}{c|}{CaOPh-CH3} & \multicolumn{2}{c|}{SrOPh-CH3} & \multicolumn{2}{c|}{CaOPh-C(CH$_3$)$_3$}& \multicolumn{2}{c|}{SrOPh-C(CH$_3$)$_3$}& \multicolumn{2}{c|}{CaOPh-Ad} & \multicolumn{2}{c}{SrOPh-Ad} \\
 \hline
Transition&$A - X$& $B-X$&$A - X$& $B-X$&$A - X$ & $B - X$&$A - X$& $B-X$&$A - X$ & $B - X$&$A - X$& $B-X$\\
 \hline
$\begin{array}{c}
     \text{0-0 Intensity}  \\
     \text{ratios } (S_0) \end{array}$
& 0.9421 & 0.9418 & 0.9556 & 0.9009 & 0.9447& 0.9383&0.9445&0.9042& 0.9293 & 0.9308 & 0.9341& 0.9200\\
\hline
$\begin{array}{c}
     \text{Curve fitting}  \\
     \text{uncertainty} (\delta S_0 )
\end{array}$  & 0.0012 & 0.0019 & 0.0005 & 0.0012&0.0012&0.0016&0.0008&0.0030 & 0.0025 & 0.0018 & 0.0011& 0.0010\\
\hline
$\begin{array}{c}
     \text{Jacobian}  \\
     \text{correction } (\Delta_J)
\end{array}$  & -0.0029 & -0.0026 & -0.0023 & -0.0044 &-0.0024 &  -0.0020 & -0.0019 &-0.0032&-0.0028 & -0.0027 & -0.0018& -0.0023\\
\hline
$\begin{array}{c}
     \text{Sensitivity}  \\
     \text{correction } (\Delta_{QE})
\end{array}$ & -0.0024 & -0.0017 & -0.0027 & -0.0049 &-0.0019 &  -0.0013  & -0.0021 &-0.0036 &-0.0046 & -0.0041 & -0.0032& -0.0038\\
\hline
$\begin{array}{c}
   \text{Uncertainty}\\
    \text{ from unobserved} \\
    \text{decays } (\delta_u)
\end{array}$& -0.0019 & -0.0006 & -0.0065 & -0.0067 & -0.0117 & -0.0011  & -0.0098 &-0.0097 & -0.0414 & -0.0227 & -0.0178 & -0.0119\\
\hline
$\begin{array}{c}
   \text{Scaled diagonal VBRs}\\
    S_0'=S_0+\Delta_J+\Delta_{QE}+\delta_u
\end{array}$  & 0.9350 & 0.9369 & 0.9442 & 0.8850 &  0.9288 &   0.9339  & 0.9329 &0.8877 & 0.8805 & 0.9015 & 0.9114 &0.9020\\
\hline\hline
\end{tabular}
    \caption{Measured intensity ratios and scaled VBRs of the diagonal 0-0 decay of all molecules. The scaling process considering contributions of unobserved vibrational decays is detailed in the section of error analysis of VBRs.
    }
    \label{DiagonalVBRs}
\end{table*}

\clearpage

\begin{figure*}
\centering
    \includegraphics[scale=0.5]{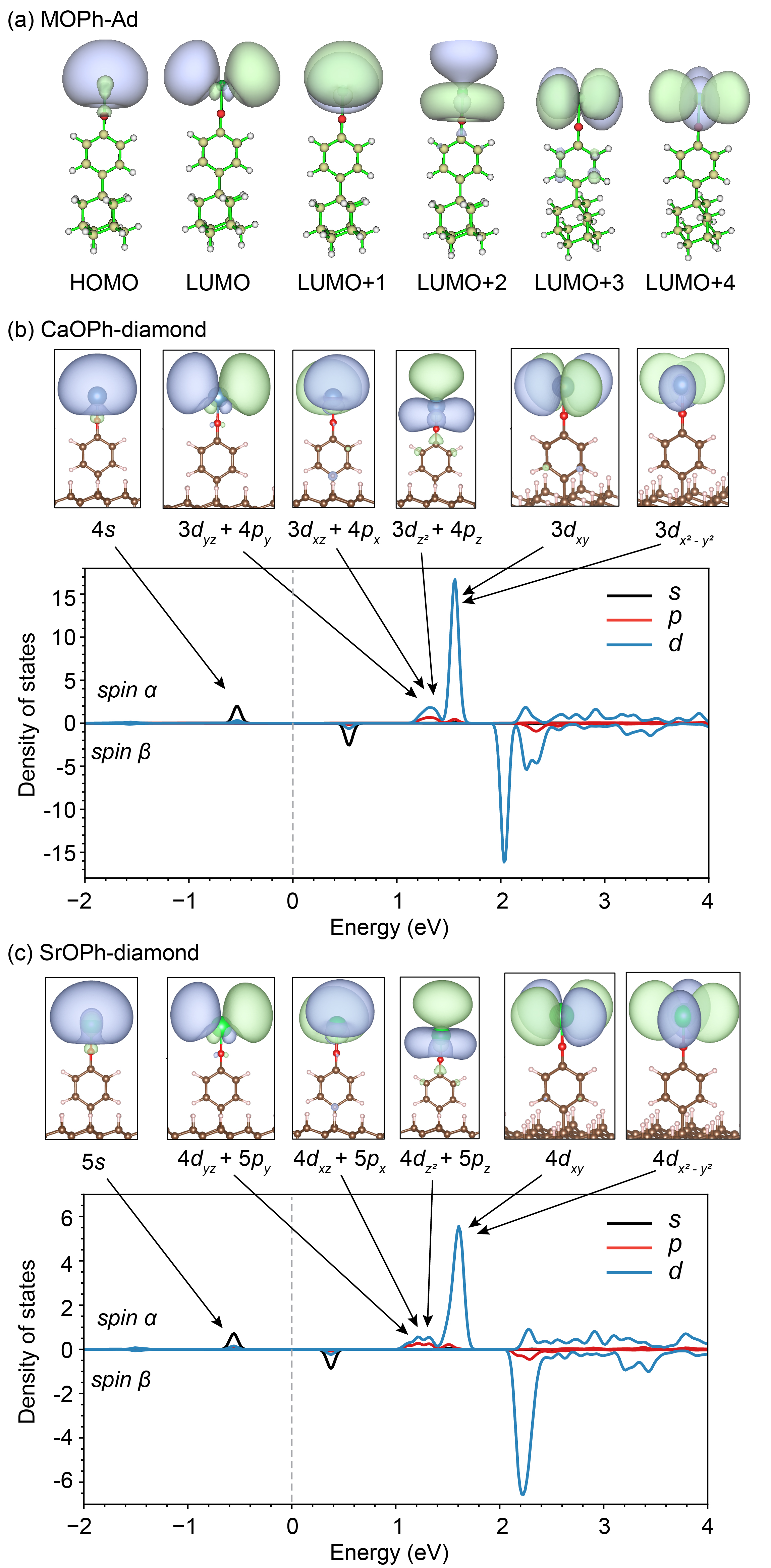}
    \caption{(a) Molecular orbitals of HOMO and LUMOs of MOPh-Ad. (b) PDOS and OCC orbitals for CaOPh-diamond. (c) PDOS and OCC orbitals for SrOPh-diamond. The PDOS plots show the contributions from \textit{s, p, d} orbitals of the metal atom.}
    \label{figs:ca-sr-pdos}
\end{figure*}

\begin{figure*}
\centering
    \includegraphics[scale=0.6]{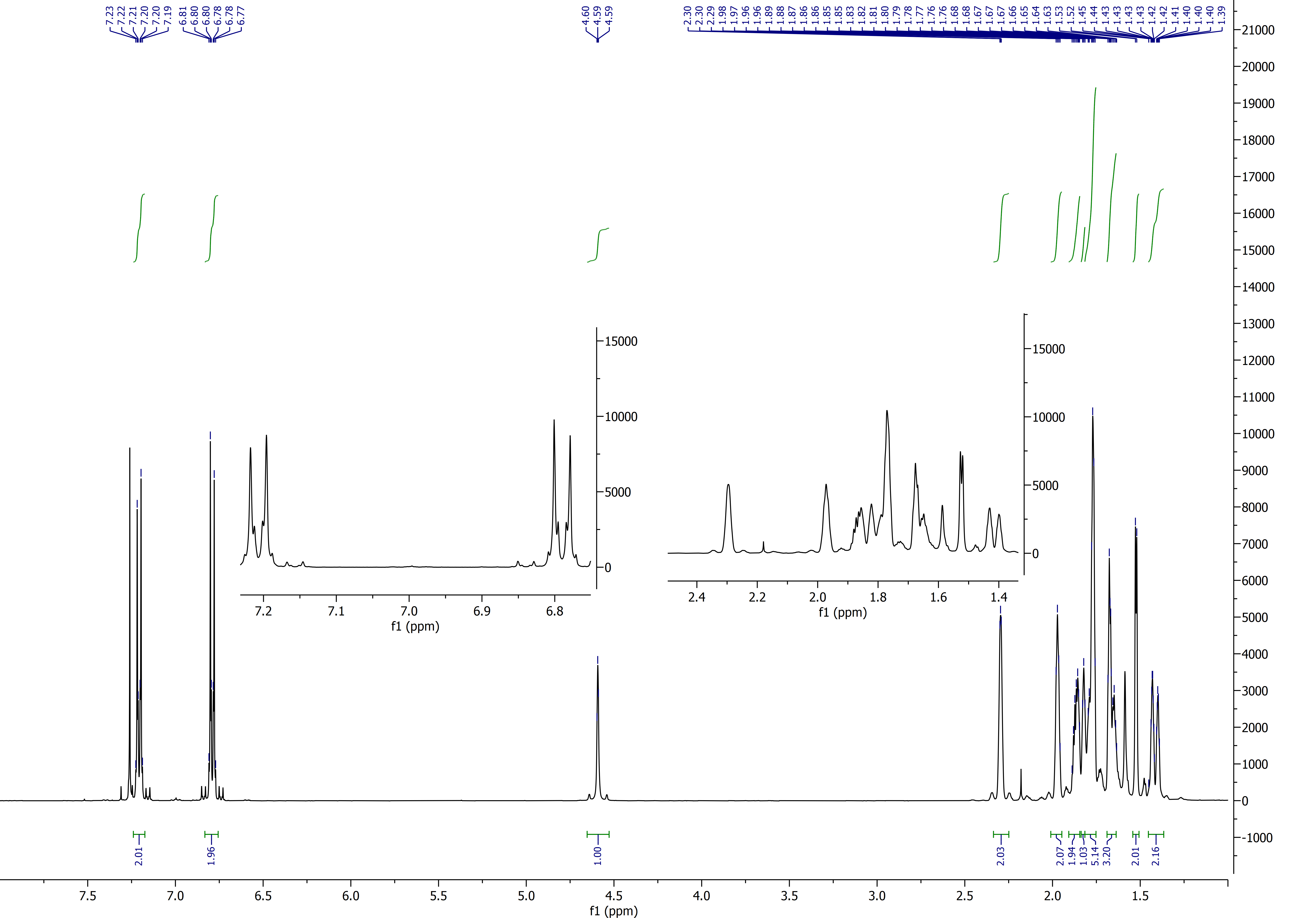}
    \caption{$^1$H NMR spectrum of 4-(1-diamantyl)-phenol. }
    \label{figs:1diA-HNMR}
\end{figure*}

\begin{figure*}
\centering
    \includegraphics[scale=0.6]{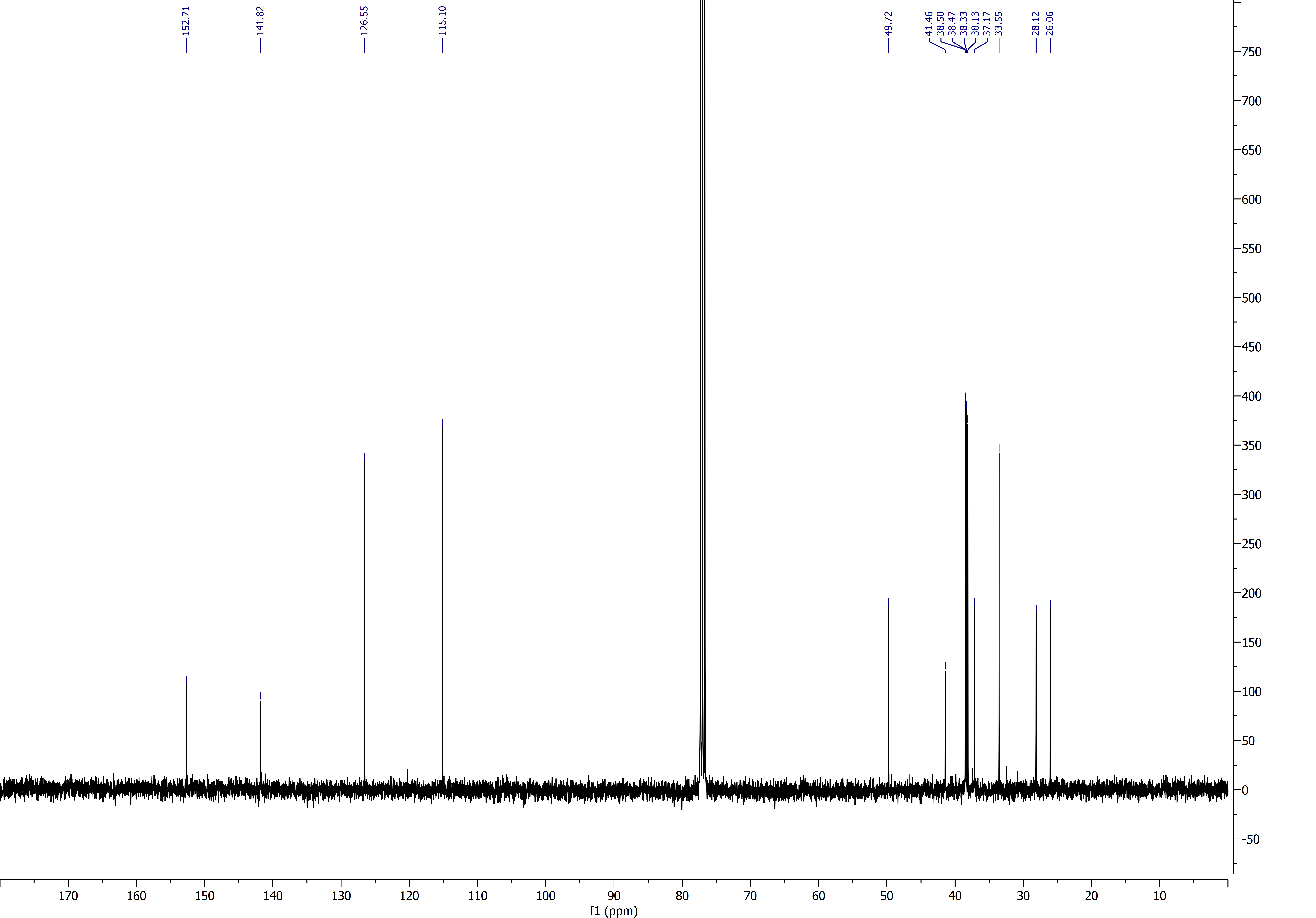}
    \caption{$^{13}$C NMR spectrum of 4-(1-diamantyl)-phenol. }
    \label{figs:1diA-CNMR}
\end{figure*}

\begin{figure*}
\centering
    \includegraphics[scale=0.6]{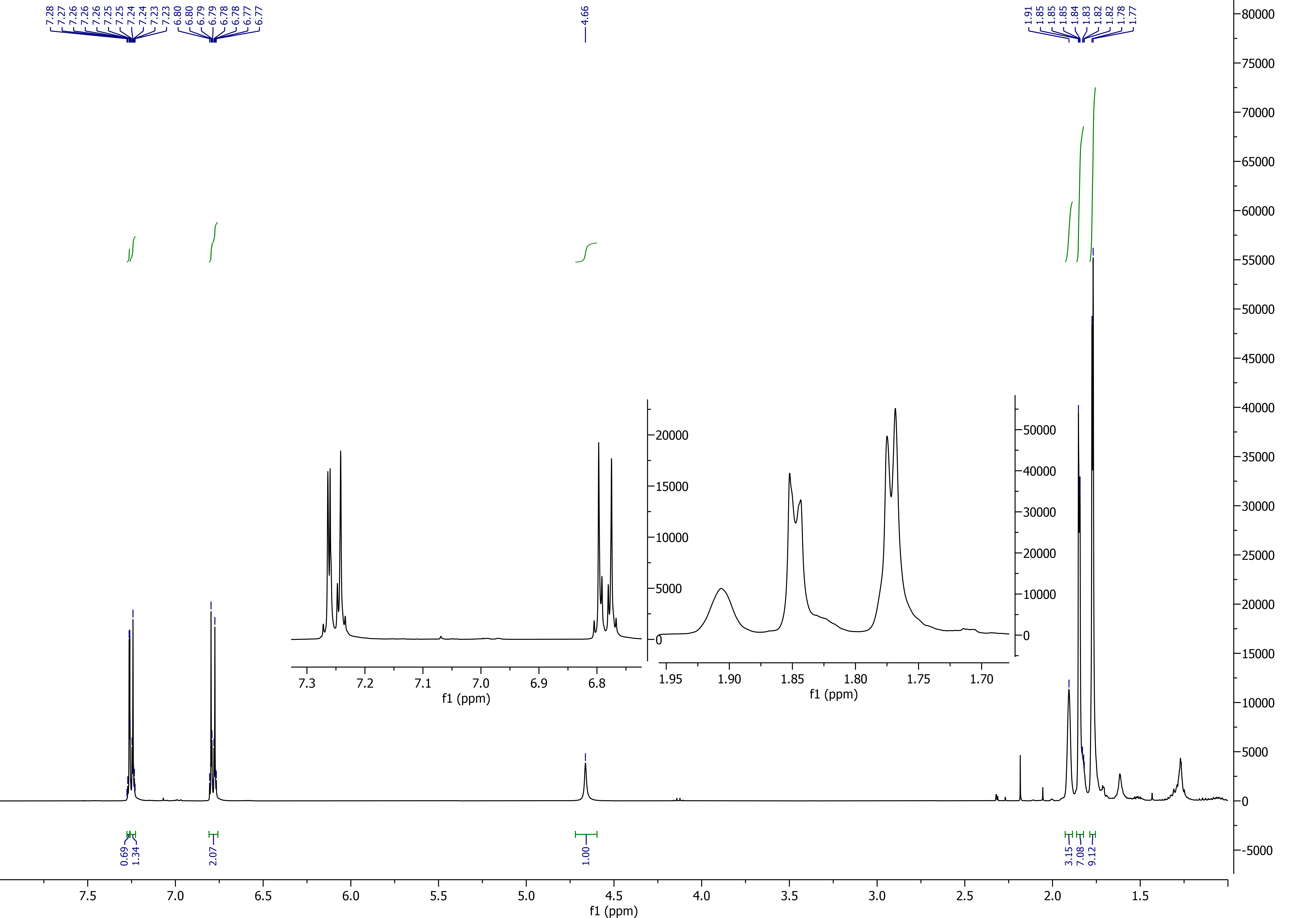}
    \caption{$^1$H NMR spectrum of 4-(4-diamantyl)-phenol. }
    \label{figs:4diA-HNMR}
\end{figure*}

\begin{figure*}
\centering
    \includegraphics[scale=0.6]{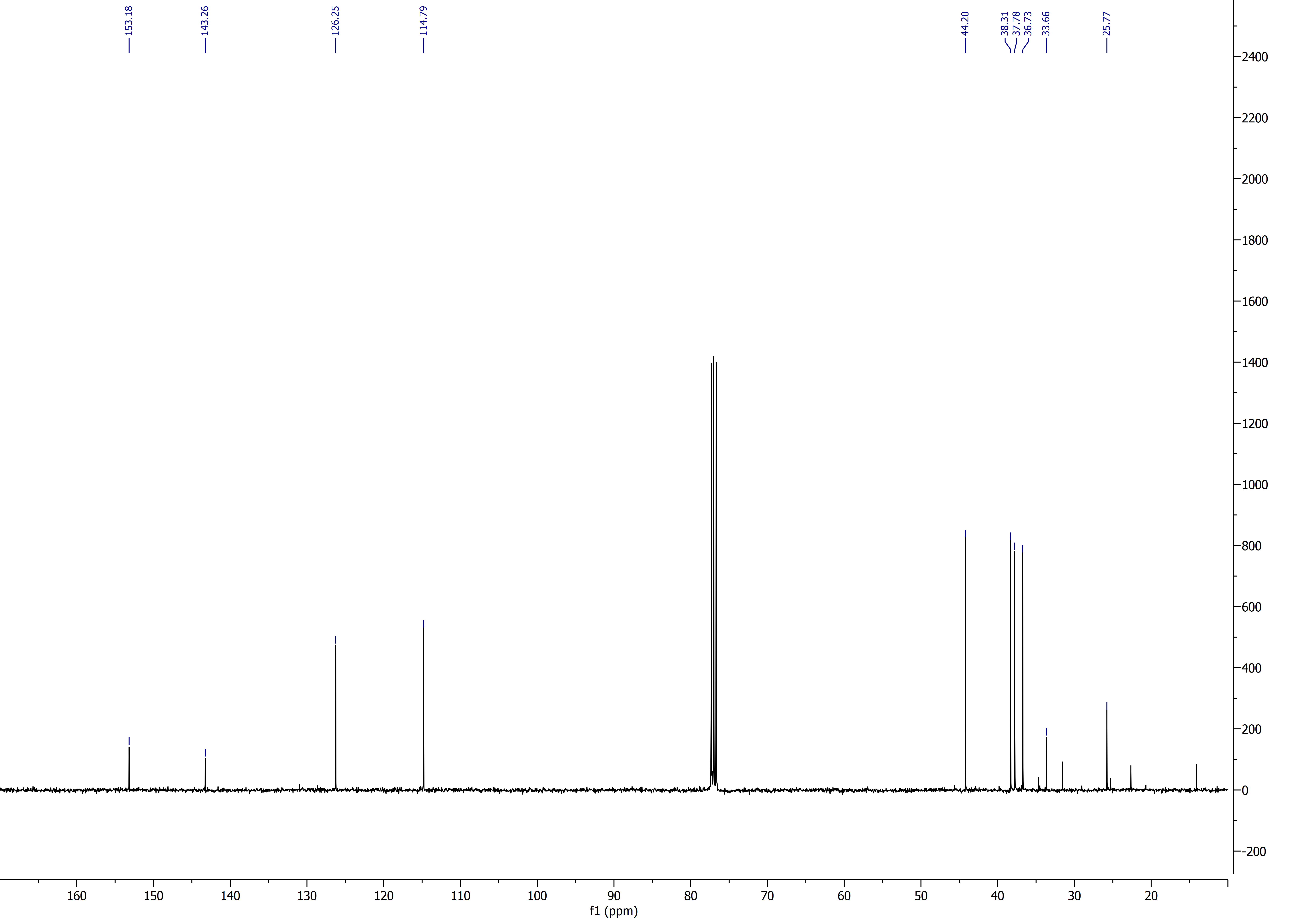}
    \caption{$^{13}$C NMR spectrum of 4-(4-diamantyl)-phenol.}
    \label{figs:4diA-CNMR}
\end{figure*}

\begin{figure*}
\centering
    \includegraphics[scale=0.75]{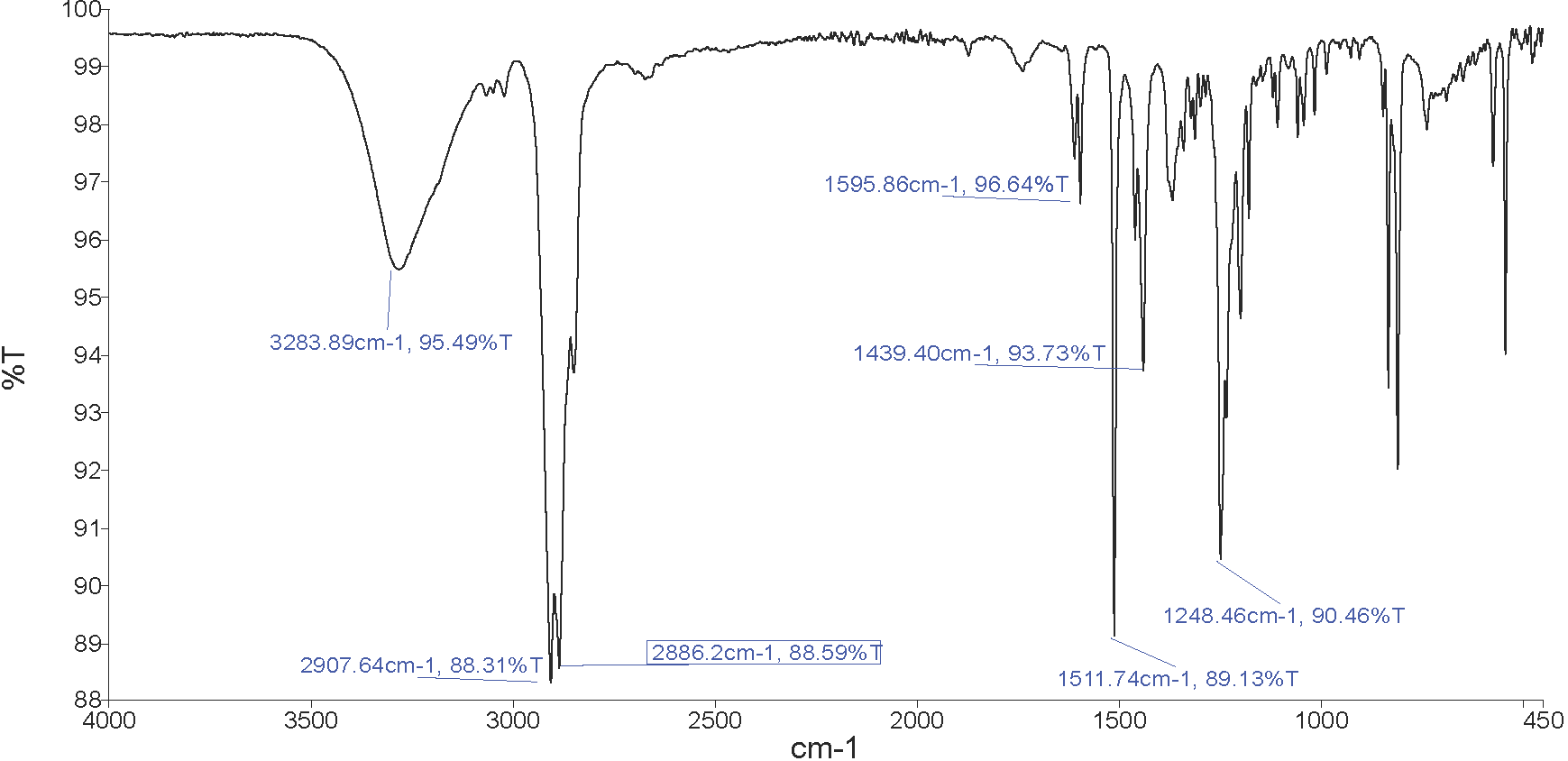}
    \caption{FT-IR spectrum of 4-(1-diamantyl)-phenol. }
    \label{figs:1diA-IR}
\end{figure*}

\begin{figure*}
\centering
    \includegraphics[scale=0.75]{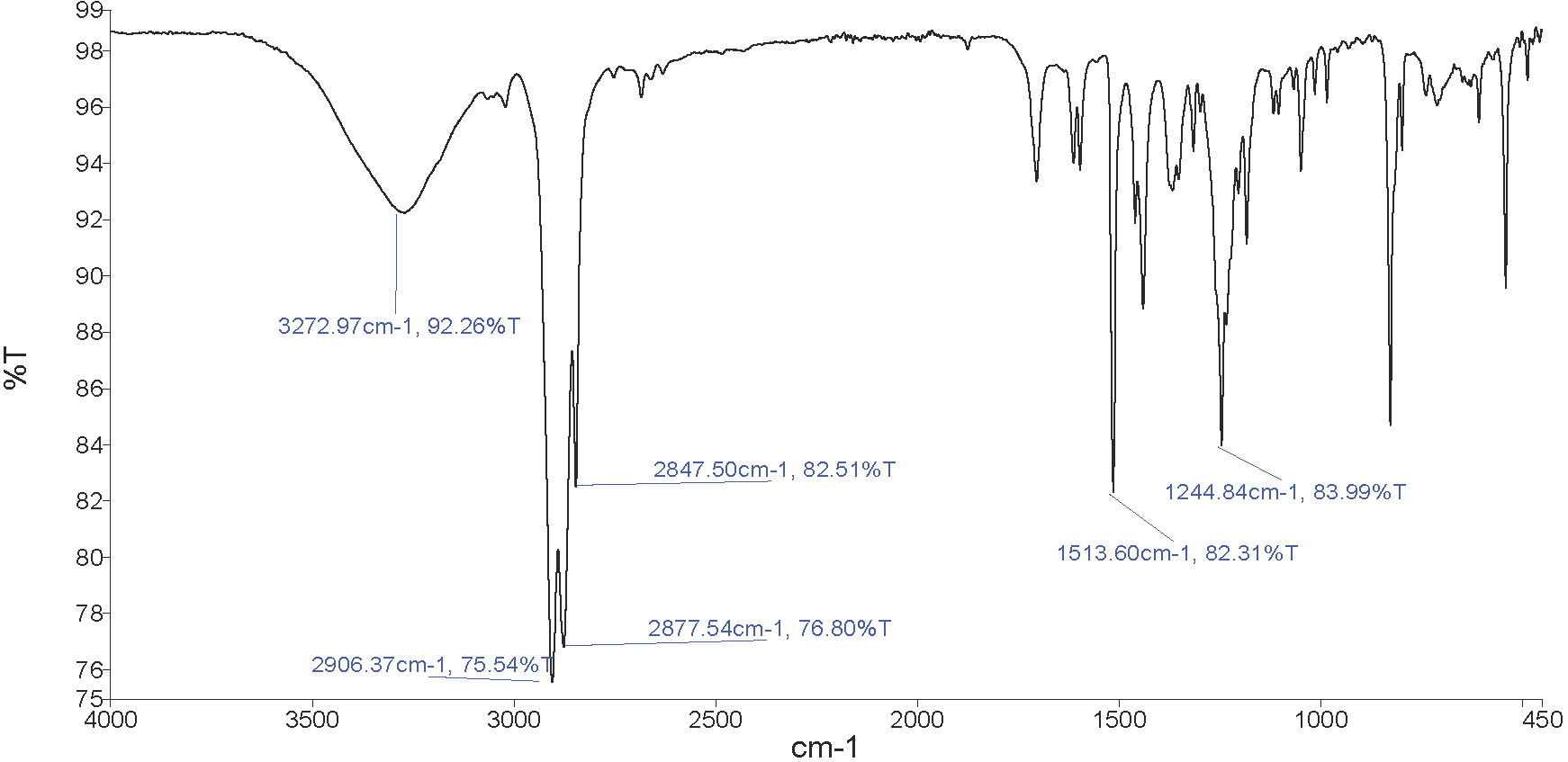}
    \caption{FT-IR spectrum of 4-(4-diamantyl)-phenol. }
    \label{figs:4diA-IR}
\end{figure*}

\end{document}